\documentclass[journal,draftcls,onecolumn,12pt,twoside]{IEEEtran}

\usepackage[utf8]{inputenc} 
\usepackage[T1]{fontenc}
\usepackage{url}
\usepackage{ifthen}
\usepackage{cite}
\usepackage[cmex10]{amsmath}
\usepackage{amsthm}
\usepackage{amssymb}
\usepackage{epsfig}
\usepackage{epstopdf}
\usepackage{graphicx}
\usepackage{float}
\usepackage{caption}
\usepackage{subcaption}
\usepackage{dsfont}
\usepackage[linesnumbered,ruled,vlined]{algorithm2e}
\usepackage[dvipsnames]{xcolor}
\usepackage{tikz}
\usepackage{pgfplots}

\usepackage{xfrac}

\usepackage{babel}
\usepackage{hyperref}
\usepackage{enumitem}
\usepackage{bbm}
\usepackage{soul}

\IEEEoverridecommandlockouts

\newtheorem{theorem}{Theorem}
\newtheorem{proposition}{Proposition}

\newtheorem{lemma}{Lemma}
\newtheorem{definition}{Definition}

\DeclareMathOperator*{\argmax}{arg\,max} \DeclareMathOperator*{\argmin}{arg\,min}

\allowdisplaybreaks

\global\long\def\s[#1]{\textnormal{\scriptsize #1}}
\global\long\def\st[#1]{\textnormal{\tiny #1}}

\global\long\def\la{\bigg(}
\global\long\def\ra{\bigg)}

\newcommand{\dfn}{\stackrel{\triangle}{=}}

\global\long\def\P{\mathbb{P}}
\global\long\def\E{\mathbb{E}}

\global\long\def\I{\mathbbm{1}}
\global\long\def\v[#1]{\mathbf{#1}} 
\global\long\def\m[#1]{\boldsymbol{#1}} 

\global\long\def\r[#1]{#1}

\global\long\def\trre[#1,#2]{\overset{{\scriptstyle (#2)}}{#1}} 

\makeatother

\newcommand{\calA}{{\cal A}}
\newcommand{\calB}{{\cal B}}
\newcommand{\calC}{{\cal C}}

\newcommand{\calF}{{\cal F}}
\newcommand{\calG}{{\cal G}}

\newcommand{\calP}{{\cal P}}

\newcommand{\calT}{{\cal T}}

\newcommand{\calX}{{\cal X}}
\newcommand{\calY}{{\cal Y}}

\newcommand{\ICC}{T}
\newcommand{\NOS}{N}
\newcommand{\dint}{\mathrm{d}}

\newcommand {\bp} {\boldsymbol{p}}

\newcommand {\bu} {\boldsymbol{u}}
\newcommand {\bv} {\boldsymbol{v}}
\newcommand {\bw} {\boldsymbol{w}}
\newcommand {\bx} {\boldsymbol{x}}
\newcommand {\by} {\boldsymbol{y}}
\newcommand {\bz} {\boldsymbol{z}}

\newcommand {\bG} {\boldsymbol{G}}

\newcommand {\bP} {\boldsymbol{P}}
\newcommand {\bQ} {\boldsymbol{Q}}
\newcommand {\bU} {\boldsymbol{U}}
\newcommand {\bV} {\boldsymbol{V}}

\newcommand {\bX} {\boldsymbol{X}}
\newcommand {\bY} {\boldsymbol{Y}}

\newcommand {\balpha} {\boldsymbol{\alpha}}

\newcommand {\nn} {\nonumber}

\begin{document}
\thispagestyle{empty}
\title{Concatenated Codes for Short-Molecule DNA Storage with Sequencing Channels of Positive Zero-Undetected-Error Capacity}

\author{Ran Tamir, Nir Weinberger and Albert Guill\'en i F\`abregas
\thanks{R. Tamir is with the Department of Signal Theory and Communications, Universitat Polit\`ecnica de Catalunya, 08034 Barcelona, Spain; email: \texttt{ran.tamir@upc.edu}. 
N. Weinberger is with the Department of Electrical and Computer Engineering, Technion, Haifa 3200003, Israel; e-mail: \texttt{nirwein@technion.ac.il}. 
Albert Guill\'en i F\`abregas is with the Department of Engineering, University
of Cambridge, CB2 1PZ Cambridge, U.K., the Department of
Signal Theory and Communications and the Institute of Mathematics (IMTech), Universitat Polit\`ecnica de Catalunya
08034 Barcelona, Spain  (e-mail: guillen@ieee.org).. 

The research of N. Weinberger was partially supported by the Israel Science Foundation (ISF), grant no. 1782/22 and the United States – Israel Binational Science Foundation (NSF-BSF), grant no. 2024763. 
The research of R. Tamir and A. Guill\'en i F\`abregas was supported in part by the European Research Council under Grants 101142747 and 101158232, and in part by the Spanish Government under Grants PID2020-116683GB-C22 and PID2021-128373OB-I00.
}}

\maketitle


\begin{abstract}
We study achievability bounds on the number of bits that can be reliably stored in a DNA-based storage system with noisy sequencing in the short-molecule regime. We analyze a concatenated coding scheme, where the outer code handles the random sampling and the inner code handles the sequencing noise. Assuming the sequencing channel is symmetric, we choose an inner code given by a linear block code with zero-undetected-error decoding. This choice simplifies the outer maximum-likelihood decoder to a tractable form, which we use to derive an achievability bound on the scaling of the number of bits that can be reliably stored.
Of independent interest, we prove that the average error probability of random linear block codes under zero-undetected-error decoding converges to zero exponentially fast with the block length, as long as the coding rate is below a critical value, known to be a lower bound on the zero-undetected-error capacity.
\end{abstract}

\begin{IEEEkeywords}
Concatenated coding, data storage, error exponents, DNA storage, linear block codes, molecular communication, permutation channel, zero-undetected-error decoding.
\end{IEEEkeywords}


\clearpage
\section{Introduction}

DNA-based data storage is characterized by its extraordinary information density\footnote{In information theory,  ``information density'' commonly denotes the random variable whose expectation is mutual information. In the context of this work, this should be understood as the number of information bits per gram of DNA.} and long-term stability, and addresses the growing demand for digital storage. A variety of working prototypes and system designs  \cite{church2012next,goldman2013towards,grass2015robust,tabatabaei2015rewritable,erlich2017dna,organick2018random,antkowiak2020low}
have catalyzed a significant body of information-theoretic and coding-theoretic
research, including coding methods \cite{sabary2024survey}, channel
capacity and error probability analysis \cite{lenz2019anchor,lenz2019coding,lenz2019upper,lenz2020achievable,lenz2020achieving,weinberger2022Error,shomorony2021dna,shomorony2022information,weinberger2022dna,ling2025exact,ling2025error,rameshwar2024information,rameshwar2025achievable,ravi2022coded,ravi2024information,narayanan2024achievable,bar2023adversarial,shomorony2021torn,mcbain2024information,mcbain2025achievable},
machine-learning based systems \cite{aharoni2025neural,welter2024end,kobovich2025input,bar2025scalable},
and secrecy \cite{vippathalla2023secure,zhang2024secret,zhang2025ramp},
among others.
In this paper, we propose and
analyze a coding scheme for DNA-based data storage with short molecules.
Our analysis focuses specifically on the scaling of the number of information bits that can be reliably stored.
To facilitate the analysis of the proposed coding scheme, we consider random, unstructured codes. Nevertheless, both the codebook ensemble and the decoder are designed with complexity constraints in mind.

\subsection{The DNA Storage Channel Model}
We study the noisy shuffling-sampling DNA storage channel model \cite{shomorony2022information}, in which the message is encoded as a multiset of $M$ molecules, each a length-$L$ string over an alphabet $\calX$ (a natural choice is ${\calX}=\{\text{A},\text{C},\text{G},\text{T}\}$ representing the four bases of DNA (Adenine, Cytosine, Guanine, and Thymine); however, for generality, we assume that $\calX$ is a general discrete set).

For a given parameter $\beta>0$, the length of the molecules is parametrized as $L=\beta\log M$. 
The $M$ molecules are gathered in a pool,
causing their original order to be completely lost. The retrieval of the stored message proceeds in two steps, repeated independently $\NOS$ times.
In the first step, a single molecule is sampled from the pool, with a uniform distribution over the $M$ molecules, and with replacement.
In the second step, which is called \emph{sequencing}, the sampled molecule is read, typically with noise, producing a length-$L$ output sequence.
The list of $\NOS$ output reads is random for two reasons: the sampling step may produce duplicates or omit molecules, and each sampled molecule is sequenced through a noisy channel.
We focus on substitution errors, although, as surveyed in \cite{heckel2019characterization}, a practical sequencing channel may also include deletions and insertions.

As explained in \cite{weinberger2022dna}, the length parameter $\beta$ affects the capacity of the storage channel, as the loss of molecule ordering has a smaller effect for larger values of $\beta$.
For example, for the DNA storage channel with ideal sampling and noiseless sequencing, if $\beta>\frac{1}{\log|{\calX}|}$,
a simple \emph{index-based} scheme achieves the channel capacity, which is given explicitly by $C=\big(\log|{\calX}|-\frac{1}{\beta}\big)^+$ \cite{shomorony2021dna}, and is monotonically increasing in $\beta$. 
For any sequencing channel, the capacity equals \emph{zero} for any $0<\beta<\frac{1}{\log|{\cal X}|}$.
However, the regime of $0<\beta<\frac{1}{\log|{\calX}|}$, which is called the \emph{short molecule regime}, is still of interest, as discussed next.

\subsection{The Short-Molecule Regime}
In the short molecule regime, as explained in \cite{gerzon2025capacity, tamir2025dna}, the information is encoded in the histogram of relative counts of each molecule type in the pool of $M$ molecules.
During the retrieval of the stored message, the sampling process produces a noisy version of this histogram; for example, molecule types having a single copy in the codeword may be sampled multiple times or not at all. The noisy sequencing process adds a further source of effective noise to the input histogram. In addition, the per-molecule sequencing channel output alphabet may differ from the input alphabet $\calX^L$, due to substitutions, deletions, or insertions.

As recently elaborated in \cite{gerzon2025capacity} and \cite{tamir2025dna},
for a given number of molecules $M$ and a given molecule length $L$, the potential total number
of reliably stored bits in the short-molecule regime may still be significant, although the capacity of the shuffling-sampling channel is zero. 
An analysis of this regime was initiated in \cite[Sec.\ 7.3]{shomorony2022information}, producing a conjecture on the maximal log-cardinality of a reliable codebook
as a function of $M$ and $L$. Specifically, \cite[Conjecture 4]{shomorony2022information}
postulates that for $\beta\in(0,\frac{1}{\log|{\calX}|})$ this log-cardinality scales asymptotically as
\begin{equation}
 \frac{1-\beta\log|{\calX}|}{2}\cdot M^{\beta\log|{\calX}|}\log M.\label{eq: optimal log-cardinality DNA}
\end{equation}

Gerzon {\em et al.} \cite{gerzon2025capacity} showed that the log-cardinality cannot exceed \eqref{eq: optimal log-cardinality DNA}, up to an $o(\text{\ensuremath{\frac{1}{\log M}}})$ additive term.
In addition, an achievability result showed that \eqref{eq: optimal log-cardinality DNA}
can be attained, but with the additional constraint that $\beta\in(\frac{1}{2\log|{\calX}|},\frac{1}{\log|{\calX}|})$,
that is, the molecules are short, but not too short.  
\cite{tamir2025dna} recently established \cite[Conjecture 4]{shomorony2022information} throughout the entire short-molecule regime $\beta\in(0,\frac{1}{\log|{\cal X}|})$.
This was achieved by conducting a random coding analysis, in which codewords are drawn by randomly choosing a point in the probability simplex based on the Dirichlet distribution and then rounding to integer count vectors.
Another contribution in \cite{tamir2025dna} is a low-complexity coding scheme termed
\emph{partition coding}.  
The codebook construction is deterministic, and decoding reduces to \emph{sorting} the frequency vector of the output reads, which standard sorting algorithms perform in $\Theta(M^{\beta\log|\calX|}\log M)$ time. This simple technique asymptotically achieves \eqref{eq: optimal log-cardinality DNA} for any $\frac{1}{3\log|\calX|} < \beta < \frac{1}{\log|\calX|}$.

\subsection{Our Contribution}
In this paper, we continue the line of work of \cite[Sec. 7.3]{shomorony2022information}, \cite{gerzon2025capacity}, and \cite{tamir2025dna}, and study the DNA storage channel with short molecules and noisy sequencing. Generalizing the analysis of \cite{tamir2025dna} to the noisy case with a general coding scheme is difficult. We therefore resort to a concatenated coding scheme, also known as a coded-index based coding scheme, in the spirit of \cite{shomorony2021dna,lenz2019upper,weinberger2022dna,lenz2020achieving,meiser2020reading,weinberger2022Error,ling2025exact}.
In the long-molecule regime, concatenated coding has been analyzed for general sequencing channels, including those with deletions and insertions \cite{weinberger2022Error,ling2024exact}. In the short molecule regime considered here, the analysis appears challenging even for general discrete memoryless sequencing channels.
We therefore restrict the family of sequencing channels and make two specific choices for the inner-coding scheme to enable a tractable analysis. We assume that the sequencing channel is symmetric, in a sense to be made precise in Section \ref{Sec_Preliminaries}.
In addition, we choose a linear block code as inner code coupled with zero-undetected-error decoding \cite{pinsker1970transmission}.
Zero-undetected-error decoding means that, given a channel output sequence, the decoder has only two possible outputs: the correct message or an erasure.
Channel symmetry, combined with the use of a linear block code, implies a \emph{message independence property}, in the spirit of \cite[Proposition 1]{hof2009performance} and \cite[Proposition 2]{hof2010performance}.
We establish this message independence property for zero-undetected-error decoding in Proposition \ref{Prop_message_indep}.
By the message independence property, and because the inner decoder never produces undetected errors, the statistics seen by the outer decoder are typically a small perturbation of those of the transmitted codeword. In this regime, the outer maximum-likelihood decoder coincides with the noiseless maximum-likelihood decoder of \cite[Eq.\ (15)]{tamir2025dna}.

Analyzing the error probability of the optimal decoder, as in the noiseless case, we prove an achievability result for the scaling law of the cardinality of the optimal storage code. This is the main result of the paper and is given in Theorem \ref{Thm_main}. For the outer decoder to attain a vanishing error probability, the inner code must achieve a vanishing erasure probability as $L\to\infty$.
We prove in Theorem \ref{THM_Linear_Performance} that the average error probability of random linear block codes under zero-undetected-error decoding converges to zero exponentially fast with the block length $L$, as long as the coding rate of the inner code is below a critical value, known to be a lower bound on the zero-undetected-error capacity. Both Proposition \ref{Prop_message_indep} and Theorem \ref{THM_Linear_Performance} provide tools for establishing results in DNA storage with noisy sequencing, but are of independent interest.

\subsection{Related Works}
The works most directly related to this study are as follows.
Motivated by DNA data storage in the short-molecule regime, \cite{gerzon2026capacity} investigated the capacity of noisy frequency-based channels, as a follow-up to \cite{gerzon2025capacity}, which provided capacity bounds for noiseless frequency-based channels. We elaborate on the connections between \cite{gerzon2026capacity} and the present paper in Section \ref{Sec_Scheme_Main_Result}.
Motivated by the short-molecule regime with Poisson sampling, \cite{bello2024lattice} studied the capacity of Poisson channels with integer (lattice) inputs.
In \cite{tamir2025achievable},
we studied frequency-based channels that allow infinite input resolution, and derived various error probability bounds.
This setup is inspired by the fact that in DNA-based storage systems, the synthesis cost scales with the number of \emph{distinct} molecule types, since once a molecule is synthesized, the cost of duplicating it is relatively low. As a consequence, any arbitrary molecule frequency vector can be accurately approximated.
In \cite{tamir2025achievable}, we also briefly discussed the connection of the short molecule regime to composite DNA storage
\cite{choi2019high,sabary2024error,zhang2025ramp,kobovich2025input,walter2024coding,walter2025coding,anavy2019data}
and to the \emph{permutation channel} \cite{kovavcevic2017coding,kovavcevic2018codes,makur2020coding,tang2023capacity,lu2024permutation}.

\subsection{Outline}
The remainder of the paper is organized as follows.
In Section \ref{Section2} we establish notation conventions, formulate the problem setting, and define the objective.
In Section \ref{Section_Motivation}, we motivate the use of a concatenated coding scheme with some specific choices.
In Section \ref{Sec_Preliminaries}, we introduce some new results for linear block codes under zero-undetected-error decoding.
In Section \ref{Sec_Scheme_Main_Result} we formulate the proposed coding scheme and present the main result.
In Section \ref{Sec_Summary} we conclude with a summary and future research directions.
The proofs of the main result and the auxiliary results are provided in the appendices.

\section{Notation Conventions and Problem Formulation} \label{Section2}
\subsection{Notation Conventions}
For a positive integer $n$, we denote $[n]=\{1,2,\ldots,n\}$. For an event $\calA$, its probability will be denoted by $\P[\calA]$ and the corresponding indicator function by $\I[\calA]$. The cardinality of a finite set $\calA$ is denoted by $|\calA|$.
The expectation of a random variable $X$ will be denoted by $\E[X]$. 
The floor function of a real number $x$ is denoted by $\lfloor x \rfloor$ and defined as $\lfloor x \rfloor = \max\{y \in \mathbb{Z}:~y \leq x\}$.
The $(n-1)$-dimensional probability simplex, denoted by $\calP_n$, is defined as
\begin{equation}
\calP_n = \bigg\{(x_1,\ldots,x_n)\in [0,1]^n:~\sum_{i=1}^{n}x_i = 1 \bigg\}.
\end{equation}
The relative entropy or Kullback--Leibler (KL) divergence between two probability mass functions (PMFs) $P$ and $Q$ on alphabet $\calX$ is defined as 
\begin{equation}
D(P\|Q) = \sum_{x\in\calX} P(x)\log \frac{P(x)}{Q(x)}.
\end{equation}
The Dirichlet distribution of order $n \geq 2$ with positive parameters $\alpha_1,\ldots,\alpha_n$ has a probability density function with respect to Lebesgue measure on the Euclidean space $\mathbb{R}^{n-1}$ given by 
\begin{align}
f(x_{1},\ldots,x_{n}) = \frac{\Gamma\la \sum_{i=1}^{n} \alpha_i \ra}{\prod_{i=1}^{n}\Gamma(\alpha_i)} \prod_{i=1}^{n} x_i^{\alpha_i-1},
\end{align}
for any $(x_{1},\ldots,x_{n}) \in \calP_n$ and zero otherwise. The gamma function is defined as
\begin{align}
\Gamma(z) = \int_{0}^{\infty} t^{z-1} e^{-t} \dint t. 
\end{align}

\subsection{Problem Formulation}
Let $\calC_M$ be a codebook for short-molecule data storage. Each codeword in $\calC_M$ is composed of at most $M$ molecules. Different codewords may have different sizes, but we impose a uniform upper bound because the input cost scales with the number of molecules synthesized.
More specifically, for any $m \in \{1,2,\ldots,|\calC_M|\}$, the codeword $\bx(m)$ is given by a set of sequences of the form
\begin{equation}
    (\bx_1^L(m), \bx_2^L(m),\ldots,\bx_{J(m)}^L(m)),
\end{equation}
where $J(m) \leq M$ and for every $i \in [J(m)]$, $\bx_i^L \in \calX^L$. 
In the short-molecule regime, we assume that for some $\beta \in (0,\frac{1}{\log|\calX|})$ 
\begin{equation}
    L = \beta \log M, 
\end{equation}
and then, the cardinality of $\calX^L$ is given by
\begin{equation}
    |\calX^L|=|\calX|^{\beta \log M}=M^{\beta\log|\calX|}.
\end{equation}

We assume that message $m$ is drawn equiprobably from the set $\{1,2,\ldots,|\calC_M|\}$ and that all the molecules that form the codeword $\bx(m)$ are placed in the molecular pool.
When the message is retrieved, we assume that exactly $\NOS$ sequences $\tilde{\bx} = (\tilde{\bx}_1^L, \tilde{\bx}_2^L,\ldots,\tilde{\bx}_\NOS^L)$ are independently sampled (with replacement) from the DNA pool.
We assume that the \textit{coverage depth} $\xi = \frac{\NOS}{M}$ is fixed. 
During sequencing, each sequence $\tilde{\bx}_{i}^{L}$, $i \in [\NOS]$, is independently corrupted by a discrete memoryless channel to produce the sequence $\by_{i}^{L} \in \calY^L$. 
Let $W=\{W(y|x):x\in\calX,y\in\calY\}$ be a probability transition matrix.
For a sampled molecule $\tilde{\bx} = (\tilde{x}_1,\ldots,\tilde{x}_L)$, the probability of observing the output vector $\by = (y_1,\ldots,y_L)$ is given by
\begin{equation}
W^{(L)}(\by|\tilde{\bx}) = \prod_{i=1}^{L}W(y_i|\tilde{x}_i).
\end{equation} 

Based on the output sequences $\by = (\by_1^L, \by_2^L,\ldots,\by_\NOS^L)$, the decoder estimates the message as $\hat{m}(\by)$.
The probability of error of any decoder is given by
\begin{align}
    \varepsilon_M = \P[ \hat{m}(\bY) \neq m ],
\end{align}
which is taken with respect to the randomness of the message selection, the sampling process, and the random sequencing noise.

Define the optimal scaling function $\Psi(M,\beta,W)$ by
\begin{equation} \label{res00}
    \sup_{\{\calC_M\}:\,\varepsilon_M \to 0}\, \limsup_{M \to \infty} \frac{\log|\calC_M|}{\Psi(M,\beta,W)} = 1.
\end{equation}
Since finding the exact $\Psi(M,\beta,W)$ is hard in general, our main objective is to prove achievability bounds on this optimal scaling function.

In the noiseless case, \cite{gerzon2025capacity, tamir2025dna} proved that
\begin{equation}
\Psi(M,\beta,W)=\frac{1-\beta\log|\calX|}{2}M^{\beta\log|\calX|}\log(M).  
\end{equation}

\section{Motivation for a Concatenated Coding Scheme} \label{Section_Motivation}      

We motivate a concatenated coding scheme: an outer code that handles the random sampling and an inner code that handles the random sequencing.

We first examine a simple, unconcatenated coding scheme, where the decoder decodes the message directly from the $\NOS$ channel outputs.
In a noisy sequencing setting,
where each sampled molecule may be transformed into a distinct molecule, the resulting maximum-likelihood decoder is relatively complicated and does not seem to lend itself to a tractable analysis.
For simplicity, we assume that the sequencing channel is a discrete memoryless channel introducing substitution errors.
Assume further that each codeword consists of exactly $M$ molecules.
Let $\{P_m(\bx)\}_{\bx\in\calX^L}$ denote the proportions of the different molecule types in the $m$th codeword.
Let $\tilde{\by}=(\by_1,\by_2,\ldots,\by_{N})$ denote the set of output samples, and for any $\by \in \calY^L$, denote the enumerators
\begin{align}
N_{\tilde{\by}}(\by) := \sum_{i=1}^{\NOS} \I\{\by_i = \by\}.
\end{align} 
In the noisy case, the likelihood is given by
\begin{align}
p\big(\by_1,\ldots,\by_\NOS|\bx(m)\big)
&= \prod_{j=1}^{\NOS} \bigg[\frac{1}{M}\sum_{i=1}^{M}W^{(L)}(\by_j|\bx_i(m))\bigg] \\
&= \prod_{j=1}^{\NOS} \bigg[\sum_{\bx\in\calX^L} W^{(L)}(\by_j|\bx)P_{m}(\bx)\bigg] \\
&= \prod_{\by\in\calY^L} \bigg[\sum_{\bx\in\calX^L} W^{(L)}(\by|\bx)P_{m}(\bx)\bigg]^{N_{\tilde{\by}}(\by)},
\end{align}
which implies that
\begin{align}\label{Decoder_noise} 
\hat{m}_{\mbox{\tiny ML}}(\tilde{\by}) =\argmax_{m} \prod_{\by\in\calY^L} \bigg[\sum_{\bx\in\calX^L} W^{(L)}(\by|\bx)P_{m}(\bx)\bigg]^{N_{\tilde{\by}}(\by)}.
\end{align}

In the noiseless case, the maximum-likelihood decoder reduces to 
\begin{align}\label{Decoder_noiseless} 
\hat{m}_{\mbox{\tiny ML}}^{\mbox{\tiny noiseless}}(\tilde{\by}) =\argmax_{m} \prod_{\bx\in\calX^L} P_{m}(\bx)^{N_{\tilde{\by}}(\bx)}.
\end{align}
\cite{tamir2025dna} proposed a random coding scheme in which each codeword is drawn from the probability simplex according to the Dirichlet distribution, and analyzed the error probability of \eqref{Decoder_noiseless} using known probabilistic results for this distribution.
However, similar methods do not apply to the generalized decoder in \eqref{Decoder_noise}, since a linear transformation of a Dirichlet-distributed random vector is, in general, no longer Dirichlet-distributed.
Motivated by the simpler form of the maximum-likelihood decoder in the noiseless case \eqref{Decoder_noiseless}, we propose a concatenated coding scheme that makes the outer decoder identical to \eqref{Decoder_noiseless}.

In the noiseless system model of \cite{tamir2025dna}, where each sampled molecule is read exactly, each codeword is drawn from all molecule types, with at most $M$ molecules in total.
Now, when the system model involves noisy sequencing, we choose a subset of all molecule types as a basis for generating the various PMF codewords such that the different molecules in this subset are relatively distant from each other. In other words, we choose an error-correcting code $\calC^{\star}$ and a suitable decoder, with the requirement that the decoder lend itself to a tractable analysis.
For some $T\in\{1,2,\ldots,|\calX|^L\}$, let the \emph{inner code} be given by $\calC^{\star}=\{\bx_1,\ldots,\bx_T\}$, where $\bx_i\in\calX^L$ for any $i\in[T]$.
Let $\mathsf{D}:\calY^L \to [T]\cup\{e\}$ be an inner decoder that assigns each sample $\by\in\calY^L$ to a message or declares an erasure.
For any $\ell \in [T]$, denote the enumerators
\begin{align}
\hat{N}_{\tilde{\by}}(\ell) := \sum_{i=1}^{\NOS} \I\{\mathsf{D}(\by_i) = \ell\}.
\end{align}

For a general inner coding scheme, the outer decoder follows the form
\begin{align} \label{Decoder_Coded} 
\hat{m}_{\mbox{\tiny ML}}(\tilde{\by}) =\argmax_{m} \prod_{\ell=1}^{\ICC} \Bigg[\frac{\sum_{k=1}^{\ICC}P_{m}(\bx_k) p(k \to \ell)}{\sum_{\ell'=1}^{\ICC}\sum_{k'=1}^{\ICC}P_{m}(\bx_{k'}) p(k' \to \ell')} \Bigg]^{\hat{N}_{\tilde{\by}}(\ell)},
\end{align}
where $p(k \to \ell)$ denotes the fraction of sampled molecules of type $k$ that are decoded as type $\ell$. The normalization stems from the fact that, for a general decoder, the vector $\{p(k\to\ell):~\ell\in[\ICC]\}$ may not be a PMF (e.g., when the decoder may declare erasures).
For the specific choice of the numbers $\{p(k \to \ell)\}$, given by
\begin{align}
\label{Req1}
p(k \to \ell) &= 0,~~\forall k,\ell \in [\ICC], k \neq \ell, \\
\label{Req2}
p(\ell \to \ell) &= \pi \in (0,1),~~\forall \ell\in[\ICC]
\end{align}
the decoder in \eqref{Decoder_Coded} reduces to the simplified form in \eqref{Decoder_noiseless}.  
To achieve \eqref{Req1} and \eqref{Req2}, we impose two requirements on the inner coding scheme:
\begin{enumerate}
    \item Undetected errors must be completely avoided, so that the proposed decoder can only decode the correct message, or output an ``erasure''. This requirement ensures that \eqref{Req1} holds.
    \item The conditional erasure probabilities when transmitting each of the codewords in $\calC^{\star}$ should be equal. This requirement ensures that \eqref{Req2} holds. 
\end{enumerate}

To satisfy the first requirement, we resort to zero-undetected-error decoding.
A zero-undetected-error decoder may abstain from a decision (declaring an erasure) whenever any decision could lead to an undetected error.
To satisfy the second requirement, we use non-binary linear block codes, for which message independence is known to hold over memoryless symmetric channels under maximum-likelihood decoding \cite{hof2009performance} and generalized (erasure/list) decoding \cite{hof2010performance}.

\section{Linear Block Codes and Zero-Undetected-Error Decoding}
\label{Sec_Preliminaries}

This section provides the preliminaries for the analysis of the DNA-based coding scheme proposed in Section \ref{Sec_Scheme_Main_Result}.
Beyond their use for the problem at hand, these results are of independent interest.

Let $\calX = \{x_0,x_1,\ldots,x_{q-1}\}$ be an alphabet with cardinality $q$. We assume an addition operation $(+)$ over the alphabet $\calX$ for which $\{\calX,+\}$ forms an Abelian group. Let $x_0=0$ be the additive identity of this group, and let $\calY$ denote the output alphabet.
Consider linear block codes over the alphabet $\calX$. Specifically, let $\bG$ be a $K \times L$ matrix with entries in $\calX$. Then, the linear block code with generator matrix $\bG$, denoted by $\calC^{\star}=\{\bx_m\}_{m=1}^{q^K}$, where $\bx_m = (x_{m,1},\ldots,x_{m,L})$, is the set of $q^K$ linear combinations of the rows of $\bG$.

\begin{definition}[Zero-undetected-error decoding] \label{ZUE_decoding}
Let $\{\bx_m\}$ be a codebook over alphabet $\calX$. The zero-undetected-error decoding rule is defined by the following decision regions:
\begin{equation} \label{Decision_Region}
    \Lambda_m = \bigg\{\by \in \calY^L~:~ W^{(L)}(\by|\bx_m)>0, \bigcap_{m'\neq m}\{W^{(L)}(\by|\bx_{m'}) = 0\} \bigg\}
\end{equation}
where $m$ is the index of the codeword. The erasure region is given by $\Lambda_{\mbox{\scriptsize er}} = \calY^L \setminus \bigcup_m\Lambda_m$.
\end{definition}
In other words, if for at least two codewords the likelihood scores are strictly positive, then the decoder outputs an erasure.  
The conditional erasure probability of the $m$th message is given by 
\begin{equation}
P_{\mbox{\scriptsize er}|m} = \sum_{\by \in \Lambda_{\mbox{\scriptsize er}}} W^{(L)}(\by|\bx_m).
\end{equation}

Before stating our message independence result, we define channel symmetry.
The following definition of channel symmetry from \cite{hof2009performance}  generalizes  the standard definition of symmetry for memoryless binary-input output-symmetric channels. 

\begin{definition}[Channel symmetry] \label{Def_channel_symmetry}
A memoryless channel characterized by a transition probability matrix $P$, an input alphabet $\calX$, and a discrete output alphabet $\calY$ is said to be \textit{symmetric} if there exists a function $\calT:\calY \times \calX \to \calY$ which satisfies the following properties:
\begin{enumerate}
    \item For every $x \in \calX$, the function $\calT(\cdot,x):\calY \to \calY$ is bijective.
    \item For every $x_1,x_2 \in \calX$ and $y \in \calY$, the following equality holds:
    \begin{equation}
        P(y|x_1) = P(\calT(y,x_2-x_1)|x_2).
    \end{equation}
\end{enumerate}
\end{definition}

A common symmetric channel is the erasure channel, which is defined by 
\begin{equation}
W(y|x)=
  \begin{cases}
    1-p  & \quad \text{for } y=x \\
    p  & \quad \text{for } y=e,
  \end{cases}
\end{equation}
for some $p\in[0,1]$. 
For this specific channel, let us choose the function $\calT(y,x)$ as  
\begin{equation}
\calT(y,x)=
  \begin{cases}
    y+x  & \quad \text{for } y \neq e \\
    e  & \quad \text{for } y=e.
  \end{cases}
\end{equation}
It is straightforward to check that $\calT(y,x)$ satisfies both requirements, which implies that the erasure channel is symmetric according to Definition \ref{Def_channel_symmetry}.   

Various message independence properties have been proved for non-binary linear block codes: with maximum-likelihood decoding in \cite{hof2009performance}, and with generalized (erasure/list) decoding in \cite{hof2010performance}.
The following result establishes that the conditional erasure probability is independent of the transmitted codeword for all memoryless symmetric channels; the proof appears in Appendix \ref{Appendix_Message_independence}.

\begin{proposition} \label{Prop_message_indep}
Let $\calC$ be a linear block code used for transmission over a memoryless and symmetric channel according to Definition \ref{Def_channel_symmetry}. 
Then, the block erasure probability, under the zero-undetected-error decoding rule in Definition \ref{ZUE_decoding}, is independent of the transmitted codeword.
\end{proposition}

Since $L$ scales logarithmically with $M$, we require $\calC^{\star}$ to have an erasure probability that vanishes exponentially in $L$ under zero-undetected-error decoding, so that it also vanishes as $M\to\infty$.
To prove the existence of such a code, we consider the ensemble of linear $(L,K)$ block codes whose generator matrix $\bG$ has $K\times L$ i.i.d.\ entries drawn equiprobably from $\calX$.

Performance bounds for binary linear block codes over binary-input output-symmetric channels were developed for maximum-likelihood decoding in \cite[Section 3.10]{viterbi2009principles}. Error exponents for typical codes from a random linear code ensemble over the binary symmetric channel were studied in \cite{barg2002random}. 
Performance bounds for non-binary linear block codes over memoryless symmetric channels were provided for maximum-likelihood decoding in \cite{hof2009performance} and for generalized (erasure/list) decoding in \cite{hof2010performance}.  
The random-coding
bound was proved to be exponentially tight for the ensemble of random linear codes at all rates  \cite{domb2015random}. 
Exponential error bounds pertaining to zero-undetected-error decoding have been derived in \cite{telatar1992multi, csiszar1995channel, ahlswede1996erasure}, with random constant composition codes.  

The following performance bound, for non-binary linear block codes over memoryless symmetric channels under zero-undetected-error decoding, is proved in Appendix \ref{appendix_proposition_proof}.

\begin{theorem} \label{THM_Linear_Performance}
Consider the ensemble of random $(L,K)$ linear block codes $\{\calC^{\star}\}$ employed for transmission over a memoryless symmetric channel with input and output alphabets $\calX$ and $\calY$, respectively. Let $P(\cdot)$ denote the uniform distribution over $\calX$, and let $W$ be the transition
probability of the channel.
Then, the average block erasure probability under the zero-undetected-error decoding rule in \eqref{Decision_Region} satisfies
\begin{equation} \label{Probability_bound}
    \E[P_{\mbox{\scriptsize er}}(\calC^{\star})] \leq \exp\Big\{-L \cdot \sup_{\rho\in(0,1]} \big(\tilde{E}_0(\rho) -\rho R \big) \Big\},
\end{equation}
where $R=\frac{K}{L}\log|\calX|$ is the code rate (in nats per channel use) and $\tilde{E}_0(\rho)$ is defined by
\begin{equation}
\tilde{E}_0(\rho) = -\log\bigg( \sum_{y\in\calY} (PW)(y) P(\calX(y))^{\rho} \bigg),
\end{equation}
where $\calX(y)$ is the set of all $x\in\calX$ for which $W(y|x)>0$.
\end{theorem}

Denote the exponent function
\begin{equation}\label{Erasure_Exponent}
\tilde{E}(R) = \sup_{\rho\in(0,1]} \big\{\tilde{E}_0(\rho) -\rho R \big\}. 
\end{equation}

The maximum attainable rate, denoted $R_{\mbox{\scriptsize max}}(W)$, is characterized in the following result, proved in Appendix \ref{appendix_proposition_attainable_rate}.

\begin{proposition} \label{Prop_attainable_rate}
    There exists a sequence of $(L,K)$ linear block codes whose average block erasure probability converges to zero exponentially fast in $L$ under zero-undetected-error decoding, as long as
    \begin{equation}
        R < R_{\mbox{\scriptsize max}}(W) = \sum_{y\in\calY} (PW)(y) \log \frac{1}{P(\calX(y))}.
        \label{R_max_expression_main}
    \end{equation}
\end{proposition}

The expression for $R_{\mbox{\scriptsize max}}(W)$, which already appeared in \cite[Eq.\ (58)]{forney1968exponential}, is known to be a lower bound on the zero-undetected-error capacity \cite[pp.\ 42-44]{telatar1992multi}, denoted by $C_{0\mbox{\scriptsize -u}}(W)$, but in some cases, like the erasure channel, it is tight. To see why this is true, observe that for the erasure channel, a uniform input distribution induces  
\begin{equation}\label{erasure_ref1}
(PW)(y)=
  \begin{cases}
    \frac{1-p}{|\calX|}  & \quad \text{for } y \in \calX \\
    p  & \quad \text{for } y=e,
  \end{cases}
\end{equation}
and, in addition,
\begin{equation}\label{erasure_ref2}
P(\calX(y))=
  \begin{cases}
    \frac{1}{|\calX|}  & \quad \text{for } y \in \calX \\
    1  & \quad \text{for } y=e.
  \end{cases}
\end{equation}
Substituting back into \eqref{R_max_expression_main}, we find that 
\begin{equation}
R_{\mbox{\scriptsize max}}(W) 
= (1-p)\log|\calX|,
\end{equation}
which is readily identified as the Shannon capacity $C(W)$ of the erasure channel, and since $C(W)$ is an upper bound on $C_{0\mbox{\scriptsize -u}}(W)$, we conclude that $R_{\mbox{\scriptsize max}}(W)$ is the optimal rate in this case. The erasure channel is one instance of a more general fact established in \cite{pinsker1970transmission}: $C_{0\mbox{\scriptsize -u}}(W)=C(W)$ for any channel whose bipartite channel graph contains no cycles.
The bipartite channel graph is the undirected bipartite graph whose two
independent sets of vertices are the input and output alphabets
of the channel, and where there is an edge between an input $x$
and an output $y$ if $W(y|x) > 0$. 

As a complement, we present another example for a parametric family of symmetric channels, for which $R_{\mbox{\scriptsize max}}(W) < C_{0\mbox{\scriptsize -u}}(W)$ for a range of parameter values. The following example was studied in detail in \cite{telatar1992multi}.
Consider the typewriter channel with alphabets $\calX=\calY=\{0,1,2\}$ and crossover probability $\epsilon \in [0,1]$. The channel transition probabilities are given by  
\begin{equation}
W(y|x)=
  \begin{cases}
    1-\epsilon  & \quad \text{for } y=x \\
    \epsilon  & \quad \text{for } (y-x)\text{ mod }3=1 \\
    0  & \quad \text{else}. 
  \end{cases}
\end{equation}

It can be shown that for any $\epsilon \in [0,1]$, $R_{\mbox{\scriptsize max}}(W) = \log\big(\frac{3}{2}\big)$. Telatar \cite[pp.\ 48-49]{telatar1992multi} established the tighter lower bound $C_{0\mbox{\scriptsize -u}}(W) \geq \log(2)-\frac{1}{2}h(\epsilon)$, where $h(t)=-t\log(t)-(1-t)\log(1-t)$ is the binary entropy function. This lower bound on $C_{0\mbox{\scriptsize -u}}(W)$ is strictly larger than $R_{\mbox{\scriptsize max}}(W)$ for any $\epsilon < 0.2622$. Moreover, for the specific value $\epsilon=\frac{1}{2}$, it follows from \cite[Theorem 1]{csiszar1995channel} that $C_{0\mbox{\scriptsize -u}}(W)=C(W)$, which equals $\log\big(\frac{3}{2}\big)$; hence $R_{\mbox{\scriptsize max}}(W) = C_{0\mbox{\scriptsize -u}}(W)$ for $\epsilon=\frac{1}{2}$.

\section{Coding Scheme and Main Result}
\label{Sec_Scheme_Main_Result}

We now describe the encoding-decoding scheme in more detail.
For a given $M$ and $\beta$, let $\ICC=\ICC(M,\beta) \in \{1,2,\ldots,|\calX|^L\}$.
We employ a linear block code $\calC^{\star}=\{\bx_1,\ldots,\bx_\ICC\}$ such that $\bx_i \in \calX^L$ for any $i \in [\ICC]$. It follows from Theorem \ref{THM_Linear_Performance} that for any $R\in(0,R_{\mbox{\scriptsize max}}(W))$, there exists an $L$-length linear block code $\calC^{\star}$ of size
\begin{equation}
\ICC 
= \exp\{LR\}
= \exp\{\beta R \log(M)\}
= M^{\beta R}
\end{equation}
with a \textit{block erasure probability} bounded above by 
\begin{equation}
    P_{\mbox{\scriptsize er}}(\calC^{\star}) 
    \leq \exp\{-L\tilde{E}(R)\}
    = \exp\{- \beta \tilde{E}(R) \log(M)\}
    = \frac{1}{M^{\beta\tilde{E}(R)}},
\end{equation} 
which converges to zero as $M\to\infty$.

Each codeword in $\calC_M$ is generated according to the following procedure. 
For message $m$, a random PMF $\bP_m = (P_m(1),\ldots,P_m(\ICC))$ is drawn from the $(\ICC-1)$-dimensional simplex $\calP_\ICC$ according to the Dirichlet distribution with vector parameters $\balpha = (1,\ldots,1)$\footnote{A simple mechanism to generate such a random PMF $(P(1),\ldots,P(\ICC))$ is as follows: draw $\ICC$ independent random variables $X_1,\ldots,X_\ICC$ from the exponential distribution with parameter $1$ and then set $P(i) = \frac{X_i}{\sum_{j=1}^{\ICC}X_j}$ for any $i \in [\ICC]$.}, which is equivalent to the uniform measure over $\calP_\ICC$.  
To turn $\bP_m$ into an empirical PMF $\hat{\bP}_m$, we choose the $m$-th codeword to contain $\lfloor M P_m(\ell) \rfloor$ copies of the string $\bx_{\ell}$, where $\ell \in [\ICC]$. The $m$-th codeword is also represented by the empirical probability vector $\hat{\bP}_m = (\hat{P}_m(1),\ldots,\hat{P}_m(\ICC))$, where for any $\ell \in [\ICC]$,  
\begin{equation}
\label{Quant_PMF}
\hat{P}_m(\ell)=\frac{\lfloor M P_m(\ell) \rfloor}{\sum_{k=1}^{\ICC} \lfloor M P_m(k) \rfloor}.     
\end{equation} 

After sampling and sequencing, the inner decoder observes $\by = (\by_1^L, \by_2^L,\ldots,\by_\NOS^L)$ and recovers each sequence in $\by$ that falls in $\bigcup_{\ell=1}^{\ICC}\Lambda_{\ell}$. The regions $\{\Lambda_\ell\}_{\ell\in[\ICC]}$ are the unambiguous decoding regions, where the decoder outputs the correct codeword.
Let $S \in \{0,1,\ldots,\NOS\}$ be the random number of recoverable sequences in $\by$.
If $S=0$, an error is declared. The probability of this error event is bounded by
\begin{align}
\P[S=0]
&= \P\Bigg[\bigcap_{i=1}^{\NOS} \bigg\{\bY_i^L \notin \bigcup_{\ell=1}^{\ICC}\Lambda_{\ell} \bigg\} \Bigg]  \\
&= \prod_{i=1}^{\NOS} \P\bigg[\bY_i^L \notin \bigcup_{\ell=1}^{\ICC}\Lambda_{\ell}\bigg]  \\
&= \prod_{i=1}^{\NOS} P_{\mbox{\scriptsize er}}(\calC^{\star}) \\
&\leq \exp\{-\tilde{E}(R) \beta \xi M\log(M)\}, 
\end{align}
which tends to zero as $M \to \infty$. 
If $S \geq 1$, the inner decoder outputs the set $\bw = (\bw_1^L, \bw_2^L,\ldots,\bw_S^L)$, where $\bw_{i}^L \in \calC^{\star}$ for any $i \in [S]$.    
By zero-undetected-error decoding, for each $i\in[S]$, $\bw_i^L$ is the \emph{only} codeword that can lead to the corresponding sequence in $\by$.

In a second step, the decoder calculates the frequency vector 
\begin{equation}
\hat{\bQ}_{\bw} = (\hat{Q}_{\bw}(1),\ldots,\hat{Q}_{\bw}(\ICC)),    
\end{equation}
where for any $\ell \in [\ICC]$, 
\begin{equation} \label{Def_Q}
    \hat{Q}_{\bw}(\ell) = \frac{1}{S}\sum_{i=1}^{S} \I[\bw_i^L = \bx_{\ell}].
\end{equation}

The outer decoder chooses the message whose codeword minimizes the KL divergence with $\hat{\bQ}_{\bw}$:
\begin{align} \label{ML_Decoder}
\hat{m}(\bw)
&=\argmin_{m \in [|\calC_M|]} D(\hat{\bQ}_{\bw}\|\hat{\bP}_m). 
\end{align}

In Appendix \ref{appendix_Main_Result} we prove the following result.
\begin{theorem} \label{Thm_main}
    Consider a noisy shuffling-sampling channel with sequencing channel $W$ of maximal rate $R_{\mbox{\scriptsize max}}(W)$, 
    molecule length parameter $\beta \in (0,\frac{1}{\log|\calX|})$, and coverage depth $\xi > 0$. 
    Then,
\begin{equation}
\Psi(M,\beta,W) \geq \frac{1-\beta R_{\mbox{\scriptsize max}}(W)}{2}M^{\beta R_{\mbox{\scriptsize max}}(W)}\log(M),
\end{equation}
where $\Psi(M,\beta,W)$ is defined in \eqref{res00}. 
\end{theorem}

A few comments are now in order:
\begin{enumerate}
    \item Although the coverage depth $\xi > 0$ is arbitrary, the asymptotic log-cardinality is independent of $\xi$. The error probability does converge faster for larger $\xi$, as also observed in \cite{weinberger2022Error, ling2025exact, ling2025error}.
    \item The symmetric-channel assumption seems difficult to relax: to our knowledge, channel symmetry is the only condition known to imply the message independence property in Proposition \ref{Prop_message_indep}.
    The simplified form of the maximum-likelihood decoder is a direct consequence of this property, and the proof of Theorem \ref{Thm_main} in Appendix \ref{appendix_Main_Result} also relies heavily on it.
    \item The proposed coding scheme is, in principle, suboptimal: decoding of the inner code (for each sampled molecule) and decoding of the outer code are performed separately, whereas \textit{joint decoding} would likely achieve a significantly lower error probability.
    \item As mentioned in the Introduction, \cite{gerzon2026capacity} recently proposed an achievability bound for the DNA channel with noisy sequencing. For a wide class of sequencing channels, \cite[Corollary 1]{gerzon2026capacity} states that
    \begin{equation}\label{Eq_Gerzon_bound}
       \Psi(M,\beta,W) \geq c(\beta,W) M^{\beta\log|\calX|}\log(M), 
    \end{equation}
    which is an improved achievability bound compared to the one in Theorem \ref{Thm_main}, since in \eqref{Eq_Gerzon_bound}, the sequencing noise only affects the leading factor, not the exponent of $M$. The result in \cite[Corollary 1]{gerzon2026capacity} holds in the range $\beta\in\big(\frac{2}{3\log|\calX|},\frac{1}{\log|\calX|}\big)$, while the result of Theorem \ref{Thm_main} holds for any $\beta\in\big(0,\frac{1}{\log|\calX|}\big)$. However, since the results in \cite{gerzon2026capacity} follow from Feinstein's maximal coding bound \cite{feinstein1954new}, \cite[Thm.\ 20.7]{polyanskiy2023information}, they may only be attainable with a scheme of much higher computational complexity than the concatenated scheme proposed here.
    \item The two papers \cite{weinberger2022Error} and \cite{ling2025exact} also consider concatenated coding schemes in the long-molecule regime. Some notable differences and similarities between our coding scheme and the one in \cite{weinberger2022Error} are as follows:
    \begin{itemize}
        \item In \cite{weinberger2022Error} the sequencing channel is general --- not necessarily memoryless or symmetric.
        In contrast to our scheme, where the inner coding scheme is restricted to a linear block code with zero-undetected-error decoding, the only requirement in \cite{weinberger2022Error} is that the inner code has a vanishing error probability as $L$ grows to infinity.
        \item Observe that both the scaling law in Theorem \ref{Thm_main} and the exponential error bounds in \cite[Theorem 3]{weinberger2022Error} do not depend on the exact performance of the inner codes, i.e., on the error probability of the inner code in \cite{weinberger2022Error} or the erasure probability of the inner code in our scheme. While \cite{weinberger2022Error} assumes that the error probability of the inner code behaves like $e^{-\Theta(L^{\zeta})}$ for some $\zeta>0$, the erasure probability of $\calC^{\star}$ does not even need to vanish --- it suffices that it not converge to one, as a consequence of the expressions in \eqref{ref_final} and \eqref{Final_Expression}.
        Nonetheless, this fact yields little improvement in the information density: for rates greater than $R_{\mbox{\scriptsize max}}(W)$, a strong converse is expected to hold, implying that the erasure probability converges exponentially fast to $1$.
        \item In our scheme, each one of the $M$ molecules of a codeword is chosen from the same inner code $\calC^{\star}$, while in \cite{weinberger2022Error}, the inner code $\calB$ is partitioned into $M$ equal cardinality sub-codes $\calB_m$, so that the $m$th molecule of a codeword is chosen only from $\calB_m$.
        This distinction stems from a fundamental difference between the long-molecule regime and the short-molecule regime. While in the short-molecule regime, each codeword is composed of many copies of each molecule type, in the long-molecule regime, all the  molecules of a codeword are distinct from one another.
        \item Since the two regimes are conceptually different, they call for different outer coding schemes. The optimal outer decoder in our coding scheme is the minimum KL divergence decoder, since in the short-molecule regime, the message is encoded in the relative frequencies of the different molecule types. As discussed above, this decoder finds the closest PMF codeword in the probability simplex. 
        The outer decoding procedure in \cite{weinberger2022Error} is different.
        After the individual molecule decoding stage, which is the same as in our scheme, the decoder holds $\NOS$ sequences from $\calB$, which are partitioned among the $M$ sub-codes $\calB_m$. For each $m\in[M]$, the decoder collects the set of inner-code decoded output molecules which belong to $\calB_m$ (if there are any), and either chooses a unique molecule from this sub-code, or declares an erasure of the $m$th molecule. To correct possible erasures or undetected erroneous molecules, the outer coding scheme in \cite{weinberger2022Error} employs a minimum Hamming distance (on a molecule level) decoder.
        \item Both the scaling law in Theorem \ref{Thm_main} and the exponential error bounds in \cite[Theorem 3]{weinberger2022Error} explicitly depend on the product $\beta R$, but the range of this product is different in both cases. In \cite{weinberger2022Error}, the inner code $\calB$ is partitioned into $M$ equal cardinality sub-codes $\calB_m$, such that the cardinality of each sub-code is given by
        \begin{equation}
        |\calB_m| = \frac{e^{RL}}{M} = M^{(\beta R-1)},
        \end{equation}
        thus it is required that $\beta R > 1$. In contrast, in the short-molecule regime, the inequality reverses, since
        \begin{equation}
            \beta R < \beta R_{\mbox{\scriptsize max}}(W) < \beta C(W) < \frac{\log|\calX|}{\log|\calX|} = 1. 
        \end{equation}
    \end{itemize}
\end{enumerate}


\section{Summary and Future Work}
\label{Sec_Summary}

We considered the information density of the DNA storage channel with noisy sequencing in the short-molecule regime.
For a symmetric sequencing channel, we designed a concatenated coding scheme in which each outer codeword is a quantization of a PMF drawn uniformly from the probability simplex, and the inner coding scheme consists of a linear block code and the zero-undetected-error decoder.
Using this scheme, we proved an achievability result on the scaling of the number of information bits that can be reliably stored.
As mentioned earlier, \cite{gerzon2026capacity} recently proved, using Feinstein's maximal coding bound, that a scaling law better than that of Theorem \ref{Thm_main} is achievable. This scaling law resembles the noiseless case, except for the leading factor. An important direction for future work is the design of coding schemes that achieve the scaling law proved in \cite{gerzon2026capacity}.

\appendices{\numberwithin{equation}{section}}

\section{Proof of Proposition \ref{Prop_message_indep}\label{Appendix_Message_independence}}

The following result, which is proved in \cite[Appendix A]{hof2009performance}, will be used in the sequel.
\begin{lemma}\label{Lemma_Auxiliary}
Let $x_1, x_2, x_3$ be arbitrary symbols in $\calX$, and let $p$ be a transition probability law of a memoryless symmetric channel. Then,
\begin{equation}
    p(\calT(\calT(y,x_1),x_2)|x_3) = p(\calT(y,x_1+x_2)|x_3),
\end{equation}
where $\calT(\cdot,\cdot)$ is a mapping which satisfies the properties in Definition \ref{Def_channel_symmetry}.
\end{lemma}

The decision regions are given by
\begin{align}
\Lambda_m 
&= \bigg\{\by \in \calY^L~:~ W^{(L)}(\by|\bx_m)>0, \bigcap_{m'\neq m} \bigg\{W^{(L)}(\by|\bx_{m'}) = 0 \bigg\}  \bigg\}  \\
\label{ref80}
&= \bigg\{\by \in \calY^L~:~ \prod_{i=1}^{L}W(y_i|x_{m,i})>0, \bigcap_{m'\neq m} \bigg\{\prod_{i=1}^{L}W(y_i|x_{m',i}) = 0\bigg\}  \bigg\}  \\
\label{ref81}
&= \bigg\{\by \in \calY^L~:~ \prod_{i=1}^{L}W(\calT(y_i,-x_{m,i})|0)>0, \bigcap_{m'\neq m}\bigg\{\prod_{i=1}^{L}W(\calT(y_i,-x_{m',i})|0) = 0\bigg\}  \bigg\} ,
\end{align}
where \eqref{ref80} holds since the channel is
memoryless and \eqref{ref81} follows from the symmetry of the channel. 
Let $\bz=(z_1,\ldots,z_L)$ be defined as
\begin{equation}
    z_i \dfn \calT(y_i,-x_{m,i}),\quad i \in \{1,\ldots,L\}, 
\end{equation}
where $m$ is the index of the transmitted codeword.
From Lemma \ref{Lemma_Auxiliary}, it follows that $\by\in\Lambda_m$ if and only if $\bz\in\tilde{\Lambda}_m$, where for any $m\in\{1,2,\ldots,q^K\}$ 
\begin{align}
\tilde{\Lambda}_m 
\dfn \bigg\{\bz \in \calY^L~:~ \prod_{i=1}^{L}W(z_i|0)>0, \bigcap_{m'\neq m}\bigg\{\prod_{i=1}^{L}W(\calT(z_i,x_{m,i}-x_{m',i})|0) = 0\bigg\}  \bigg\}.
\end{align}

Using the linearity of the code, it follows that 
\begin{align}
\tilde{\Lambda}_m 
= \bigg\{\bz \in \calY^L~:~ \prod_{i=1}^{L}W(z_i|0)>0, \bigcap_{\ell\neq 0}\bigg\{\prod_{i=1}^{L}W(\calT(z_i,x_{\ell,i})|0) = 0\bigg\}  \bigg\}.
\end{align}
Since the set $\tilde{\Lambda}_m$ is independent of the index $m$, we have $\tilde{\Lambda}_m=\tilde{\Lambda}_1$ for all $m \in \{1,\ldots,q^K\}$.

As a result, the conditional correct decoding probability of the $m$th message satisfies 
\begin{align}
P_{\mbox{\scriptsize c}|m} 
&= \sum_{\by \in \Lambda_m} W^{(L)}(\by|\bx_m) \\
&= \sum_{\bz \in \tilde{\Lambda}_m} W^{(L)}(\bz|\boldsymbol{0}) \\
&= \sum_{\bz \in \tilde{\Lambda}_1} W^{(L)}(\bz|\boldsymbol{0}).
\end{align}

This concludes the proof.

\section{Proof of Theorem \ref{THM_Linear_Performance}
\label{appendix_proposition_proof}}

Let $P^{(L)}(\cdot)$ denote the uniform distribution over $\calX^L$.
For a given $\by\in\calY^L$, let $\calX^L(\by)$ denote the set of all $\bx\in\calX^L$ for which $W^{(L)}(\by|\bx)>0$.
We will use below the property that $\calX^L(\by)$ is a product set. This property holds since   
\begin{align}
\calX^L(\by)
&= \{\bx\in\calX^L:~W^{(L)}(\by|\bx)>0\} \\
&= \bigg\{\bx\in\calX^L:~ \prod_{i=1}^{L}W(y_i|x_i)>0\bigg\} \\
&= \bigg\{\bx\in\calX^L:~ \bigcap_{i=1}^{L}\{W(y_i|x_i)>0\}\bigg\} \\
&= \prod_{i=1}^{L} \{x_i\in\calX: W(y_i|x_i)>0\} \\
&= \prod_{i=1}^{L} \calX(y_i).
\end{align}

By the message independence property in Proposition \ref{Prop_message_indep}, we assume without loss of generality that the encoded message is $m=1$, i.e., the all-zero sequence is transmitted over the channel.
Given the channel output sequence $\bY=\by$, the conditional probability of erasure is bounded as
\begin{align}
P_{\mbox{\scriptsize er}}(\by)
&= \P\bigg[\bigcup_{m=2}^{e^{LR}} \{W^{(L)}(\by|\bX_m)>0\} \bigg] \\
\label{ref90}
&\leq \bigg(\sum_{m=2}^{e^{LR}} \P[ W^{(L)}(\by|\bX_m)>0 ]\bigg)^{\rho} \\
\label{ref91}
&= \bigg(\sum_{m=2}^{e^{LR}} P^{(L)}(\calX^L(\by)) \bigg)^{\rho} \\
&\leq e^{\rho LR} P^{(L)}(\calX^L(\by))^{\rho},
\end{align}
where \eqref{ref90} holds for any $\rho\in(0,1]$, and in \eqref{ref91}, we used the fact that the marginal distribution of each of the codewords $\{\bX_2,\ldots,\bX_{e^{LR}}\}$ is uniform over $\calX^L$.

Averaging with respect to the channel output, we arrive at
\begin{align} \label{Intermediate_bound}
P_{\mbox{\scriptsize er}}
&\leq e^{\rho LR} \sum_{\by\in\calY^L} W^{(L)}(\by|\boldsymbol{0}) P^{(L)}(\calX^L(\by))^{\rho}.
\end{align}

For any $\bx=(x_1,\ldots,x_L)\in\calX^L$, we have that
\begin{align}
&\sum_{\by\in\calY^L} W^{(L)}(\by|\bx) P^{(L)}(\calX^L(\by))^{\rho} \nn \\
\label{ref98}
&~~~~= \sum_{\by\in\calY^L} \prod_{i=1}^{L}W(y_i|x_i) \bigg(\prod_{i=1}^{L} \frac{|\{x:W(y_i|x)>0\}|}{|\calX|}\bigg)^{\rho} \\
\label{ref95}
&~~~~= \sum_{\by\in\calY^L} \prod_{i=1}^{L}W(\calT(y_i,-x_i)|0) \bigg(\prod_{i=1}^{L} \frac{|\{x:W(\calT(y_i,-x)|0)>0\}|}{|\calX|}\bigg)^{\rho} \\
&~~~~= \sum_{\by\in\calY^L} \prod_{i=1}^{L}W(\calT(y_i,-x_i)|0) \bigg(\prod_{i=1}^{L} \frac{|\{x:W(\calT(y_i,-x_i+x_i-x)|0)>0\}|}{|\calX|}\bigg)^{\rho} \\
\label{ref92}
&~~~~= \sum_{\by\in\calY^L} \prod_{i=1}^{L}W(\calT(y_i,-x_i)|0) \bigg(\prod_{i=1}^{L} \frac{|\{x:W(\calT(\calT(y_i,-x_i),x_i-x)|0)>0\}|}{|\calX|}\bigg)^{\rho},
\end{align}
where \eqref{ref98} follows since $\calX^L(\by)$ is a product set and $P^{(L)}$ is a product distribution, \eqref{ref95} is due to the channel symmetry and \eqref{ref92} follows from Lemma \ref{Lemma_Auxiliary}.
Let $\tilde{\by}=(\tilde{y}_1,\ldots,\tilde{y}_L)$ be defined as 
\begin{equation}
\tilde{y}_i \dfn \calT(y_i,-x_i),\quad i \in \{1,\ldots,L\}, 
\end{equation}
and then
\begin{align}
&\sum_{\by\in\calY^L} W^{(L)}(\by|\bx) P^{(L)}(\calX^L(\by))^{\rho} \nn \\
&~~~~= \sum_{\tilde{\by}\in\calY^L} \prod_{i=1}^{L}W(\tilde{y}_i|0) \bigg(\prod_{i=1}^{L} \frac{|\{x:W(\calT(\tilde{y}_i,x_i-x)|0)>0\}|}{|\calX|}\bigg)^{\rho} \\
\label{ref96}
&~~~~= \sum_{\tilde{\by}\in\calY^L} \prod_{i=1}^{L}W(\tilde{y}_i|0) \bigg(\prod_{i=1}^{L} \frac{|\{x:W(\tilde{y}_i|x-x_i)>0\}|}{|\calX|}\bigg)^{\rho} \\
\label{ref97}
&~~~~= \sum_{\tilde{\by}\in\calY^L} \prod_{i=1}^{L}W(\tilde{y}_i|0) \bigg(\prod_{i=1}^{L} \frac{|\{x:W(\tilde{y}_i|x)>0\}|}{|\calX|}\bigg)^{\rho} \\
&~~~~= \sum_{\tilde{\by}\in\calY^L} W^{(L)}(\tilde{\by}|\boldsymbol{0}) P^{(L)}(\calX^L(\tilde{\by}))^{\rho},
\end{align}
where \eqref{ref96} and \eqref{ref97} are due to the channel symmetry.

It follows that for any $\bx\in\calX^L$
\begin{equation}
\sum_{\by\in\calY^L} W^{(L)}(\by|\boldsymbol{0}) P^{(L)}(\calX^L(\by))^{\rho}
= \sum_{\by\in\calY^L} W^{(L)}(\by|\bx) P^{(L)}(\calX^L(\by))^{\rho},
\end{equation}
which implies from \eqref{Intermediate_bound} that for any $\bx\in\calX^L$
\begin{align}
P_{\mbox{\scriptsize er}}
\label{ref93}
&\leq e^{\rho LR} \sum_{\by\in\calY^L} W^{(L)}(\by|\bx) P^{(L)}(\calX^L(\by))^{\rho},
\end{align}
and since only the right-hand side of \eqref{ref93} depends on $\bx$, 
averaging with respect to $P^L$ yields that 
\begin{align}
P_{\mbox{\scriptsize er}}
&\leq e^{\rho LR} \sum_{\bx\in\calX^L} P^{(L)}(\bx) \sum_{\by\in\calY^L} W^{(L)}(\by|\bx) P^{(L)}(\calX^L(\by))^{\rho} \\
&= e^{\rho LR} \sum_{\by\in\calY^L} \bigg(\sum_{\bx\in\calX^L} P^{(L)}(\bx) W^{(L)}(\by|\bx)\bigg) P^{(L)}(\calX^L(\by))^{\rho} \\
&= e^{\rho LR} \sum_{\by\in\calY^L} (P^{(L)}W^{(L)})(\by) P^{(L)}(\calX^L(\by))^{\rho} \\
\label{ref94}
&= e^{\rho LR} \bigg(\sum_{y\in\calY} (PW)(y) P(\calX(y))^{\rho}\bigg)^L,
\end{align}
where the factorization in \eqref{ref94} follows from the fact that $P^{(L)}(\cdot)$ is the uniform distribution over $\calX^L$, which implies that $P^{(L)}(\cdot)$ is a product distribution, and the fact that $\calX^L(\by)$ is a product set.
More explicitly, \eqref{ref94} is justified since for any $\by=(y_1,\ldots,y_L)$,
\begin{equation}
    (P^{(L)}W^{(L)})(\by) = \prod_{i=1}^{L} (PW)(y_i),
\end{equation}
and furthermore,
\begin{equation}
P^{(L)}(\calX^L(\by)) 
= \prod_{i=1}^{L} P(\calX(y_i)).
\end{equation}

Finally, \eqref{ref94} implies that 
\begin{align}
P_{\mbox{\scriptsize er}}
\leq \exp\bigg\{-L \bigg[-\log\bigg( \sum_{y\in\calY} (PW)(y) P(\calX(y))^{\rho} \bigg) -\rho R \bigg] \bigg\},
\end{align}
and the proof of Theorem \ref{THM_Linear_Performance} is completed by maximizing the exponent function over $\rho\in(0,1]$.

\section{Proof of Proposition \ref{Prop_attainable_rate}
\label{appendix_proposition_attainable_rate}}

It follows from Theorem \ref{THM_Linear_Performance} that any rate $R$ for which 
\begin{equation}
R < \frac{\tilde{E}_0(\rho)}{\rho} = -\log\bigg(\E[P(\calX(Y))^{\rho}]\bigg)^{\frac{1}{\rho}}
\end{equation}
for some $\rho\in(0,1]$,
is an attainable coding rate for zero-undetected-error coding with linear codes.
To find the maximum attainable rate, define the function
\begin{equation}
f(\rho) = \log(\E[X^{\rho}])^{\frac{1}{\rho}}
\end{equation}
and prove that it is monotonically non-decreasing for any $\rho>0$. To this end, for any $\rho_1 \leq \rho_2$, it follows from Jensen's inequality that
\begin{align}
f(\rho_1)
&= \log(\E[X^{\rho_1}])^{\frac{1}{\rho_1}} \\
&= \log\bigg(\E\bigg[X^{\rho_2 \cdot \frac{\rho_1}{\rho_2}}\bigg]\bigg)^{\frac{1}{\rho_1}} \\
&\leq \log(\E[X^{\rho_2}])^{ \frac{\rho_1}{\rho_2} \cdot \frac{1}{\rho_1}} \\
&= \log(\E[X^{\rho_2}])^{ \frac{1}{\rho_2}} \\
&= f(\rho_2).
\end{align}

Hence, the function $\frac{\tilde{E}_0(\rho)}{\rho}$ is monotonically non-increasing for any $\rho>0$ and the maximal attainable rate is given by
\begin{align}
R_{\mbox{\scriptsize max}}(W)
&= \sup_{\rho\in(0,1]} \frac{\tilde{E}_0(\rho)}{\rho} \\
&= \lim_{\rho \to 0} \frac{\tilde{E}_0(\rho)}{\rho} \\
\label{R_max_expression}
&= \sum_{y\in\calY} (PW)(y) \log \frac{1}{P(\calX(y))},
\end{align}
which follows from L'Hospital's rule.
The proof of Proposition \ref{Prop_attainable_rate} is complete.

\section{Proof of Theorem \ref{Thm_main}
\label{appendix_Main_Result}}

Following the inner (linear) code proof of existence in Section \ref{Sec_Preliminaries}, let $R\in(0,R_{\mbox{\scriptsize max}}(W))$ be fixed, such that there exists an $L$-length linear block code $\calC^{\star}$ of size
\begin{equation} \label{Def_N}
\ICC = M^{\beta R}
\end{equation}
for some $\beta\in\bigg(0,\frac{1}{\log|\calX|}\bigg)$, and an erasure probability $P_{\mbox{\scriptsize er}}(\calC^{\star})$ converging to zero as $M \to \infty$.

We assume without loss of generality that the encoded message is $m=1$. 
Conditioned on the transmitted codeword $\hat{\bP}_1=\hat{\bp}=(\hat{p}_1,\ldots,\hat{p}_\ICC)$, let $\bU = (U_1,\ldots,U_\ICC) \sim \text{Multinomial}(\NOS,\hat{\bp})$ denote the number of samples collected of each molecule type.
Let $\bV = (V_1,\ldots,V_\ICC)$ be the number of samples remaining after inner decoding, and let $S = \sum_{i=1}^{\ICC}V_i$.
Given $\bU=\bu=(u_1,\ldots,u_\ICC)$, we have that $V_i \sim \text{Bin}(u_i,1-P_{\mbox{\scriptsize er}}(\calC^{\star}))$, $i\in[\ICC]$, since the conditional erasure probabilities are independent of the transmitted codeword. 

For a given $\bV=\bv=(v_1,\ldots,v_\ICC)$ and $S=s$, denote the frequency vector
\begin{equation}
\hat{\bQ}_{\bv} = (\hat{Q}_{\bv}(1),\ldots,\hat{Q}_{\bv}(\ICC)),    
\end{equation}
where for any $i \in [\ICC]$, 
\begin{equation} \label{Def_Q_Alternative}
    \hat{Q}_{\bv}(i) = \frac{v_i}{s}.
\end{equation}

We denote the competing codewords $\hat{\bP}_m=(\hat{P}_m(1),\ldots,\hat{P}_m(\ICC))$, where $m \in \{2,\ldots,|\calC_M|\}$.
Given $\hat{\bP}_1=\hat{\bp}$, $\bU=\bu$, and $\bV=\bv$, the conditional probability of error is given by 
\begin{align}
\varepsilon_M(\hat{\bp},\bu,\bv)
&= \P\bigg[ \bigcup_{m=2}^{|\calC_M|} \{D(\hat{\bQ}_{\bv}\|\hat{\bP}_m) \leq D(\hat{\bQ}_{\bv}\|\hat{\bp}) \} \bigg] \\
\label{ref00}
&\leq \min \bigg\{ 1, \sum_{m=2}^{|\calC_M|} \P[D(\hat{\bQ}_{\bv}\|\hat{\bP}_m) \leq D(\hat{\bQ}_{\bv}\|\hat{\bp})] \bigg\}, 
\end{align}
using the clipped union bound, where the pairwise error probability $\P[D(\hat{\bQ}_{\bv}\|\hat{\bP}_m) \leq D(\hat{\bQ}_{\bv}\|\hat{\bp})]$ is the probability of deciding in favor of message $m$ when message $1$ was sent for a fixed $\bv$.

Let $\theta \geq 0$ be an arbitrary parameter.
The probability in \eqref{ref00} is given by
\begin{align}
\P[D(\hat{\bQ}_{\bv}\|\hat{\bP}_m) \leq D(\hat{\bQ}_{\bv}\|\hat{\bp})]
&= \P\bigg[ \sum_{i=1}^{\ICC} \hat{Q}_{\bv}(i) \log \frac{\hat{Q}_{\bv}(i)}{\hat{P}_m(i)} \leq \sum_{i=1}^{\ICC} \hat{Q}_{\bv}(i) \log \frac{\hat{Q}_{\bv}(i)}{\hat{p}_i} \bigg] \\
&= \P\bigg[ \sum_{i=1}^{\ICC} \hat{Q}_{\bv}(i) \log \hat{P}_m(i) \geq \sum_{i=1}^{\ICC} \hat{Q}_{\bv}(i) \log \hat{p}_i \bigg] \\
&= \P\bigg[ \sum_{i=1}^{\ICC} \log \hat{P}_m(i)^{\theta\hat{Q}_{\bv}(i)} \geq \theta \sum_{i=1}^{\ICC} \hat{Q}_{\bv}(i) \log \hat{p}_i \bigg] \\
&= \P\bigg[ \prod_{i=1}^{\ICC} \hat{P}_m(i)^{\theta\hat{Q}_{\bv}(i)} \geq \exp \bigg\{ \theta \sum_{i=1}^{\ICC} \hat{Q}_{\bv}(i) \log \hat{p}_i \bigg\} \bigg] \\
\label{ref0}
&\leq \frac{\E\bigg[ \prod_{i=1}^{\ICC} \hat{P}_m(i)^{\theta\hat{Q}_{\bv}(i)} \bigg]}{\exp \bigg\{ \theta \sum_{i=1}^{\ICC} \hat{Q}_{\bv}(i) \log \hat{p}_i \bigg\}},
\end{align}
where \eqref{ref0} follows from Markov's inequality. 

We upper-bound the empirical probabilities $\{\hat{P}_m(i)\}$ defined in \eqref{Quant_PMF} as
\begin{align}
\hat{P}_m(i)
&= \frac{\lfloor MP_m(i) \rfloor}{\sum_{k=1}^{\ICC} \lfloor MP_m(k) \rfloor} \\
&\leq \frac{MP_m(i)}{\sum_{k=1}^{\ICC} ( MP_m(k) - 1 )} \\
\label{Prob_upper_bound}
&= \frac{MP_m(i)}{M-\ICC}.
\end{align}
We then bound the expectation in \eqref{ref0} as
\begin{align}
\E\bigg[ \prod_{i=1}^{\ICC} \hat{P}_m(i)^{\theta\hat{Q}_{\bv}(i)} \bigg]
\label{ref1zzz}
&\leq \E\bigg[ \prod_{i=1}^{\ICC} \bigg( \frac{MP_m(i)}{M-\ICC} \bigg)^{\theta\hat{Q}_{\bv}(i)} \bigg] \\
\label{ref1}
&= \bigg(\frac{M}{M-\ICC}\bigg)^{\theta} \cdot
\E\bigg[ \prod_{i=1}^{\ICC} P_{m}(i)^{\theta\hat{Q}_{\bv}(i)} \bigg].
\end{align}

In order to evaluate the expectation in \eqref{ref1}, we use \cite[Proposition 1]{tamir2025dna} with $\alpha_1=\ldots=\alpha_\ICC=1$ and $\beta_i = \theta\hat{Q}_{\bv}(i)$, giving 
\begin{align}
\E\bigg[\prod_{i=1}^{\ICC} P_m(i)^{\theta\hat{Q}_{\bv}(i)} \bigg]
&= \frac{\Gamma(\ICC)}{\Gamma\bigg( \sum_{i=1}^{\ICC} (1+\theta\hat{Q}_{\bv}(i)) \bigg)} \cdot
\prod_{i=1}^{\ICC} \frac{\Gamma\bigg( 1+\theta\hat{Q}_{\bv}(i) \bigg)}{\Gamma(1)} \\
&= \frac{\Gamma(\ICC)}{\Gamma(\ICC+\theta)} \cdot
\prod_{i=1}^{\ICC} \Gamma\bigg( 1+\theta\hat{Q}_{\bv}(i) \bigg),
\end{align}
since $\Gamma(1)=1$.

Substituting back into \eqref{ref1} and then into \eqref{ref0}, we arrive at
\begin{align}
&\P[D(\hat{\bQ}_{\bv}\|\hat{\bP}_m) \leq D(\hat{\bQ}_{\bv}\|\hat{\bp})] \nn \\
&~~\leq 
\bigg(\frac{M}{M-\ICC}\bigg)^{\theta} \cdot
\frac{\Gamma(\ICC)}{\Gamma(\ICC+\theta)} \cdot
\bigg( \prod_{i=1}^{\ICC} \Gamma\bigg( 1+\theta\hat{Q}_{\bv}(i) \bigg) \bigg)
\cdot \exp \bigg\{ -\theta \sum_{i=1}^{\ICC} \hat{Q}_{\bv}(i) \log \hat{p}_i \bigg\}.
\end{align}

Since the bound is valid for any $\theta \geq 0$, we choose $\theta=s$, which results in
\begin{align}
&\P[D(\hat{\bQ}_{\bv}\|\hat{\bP}_m) \leq D(\hat{\bQ}_{\bv}\|\hat{\bp})] \nn \\
\label{ref2}
&~~\leq 
\bigg(\frac{M}{M-\ICC}\bigg)^{s} \cdot
\frac{\Gamma(\ICC)}{\Gamma(\ICC+s)} \cdot
\bigg( \prod_{i=1}^{\ICC} \Gamma\bigg( 1+s\hat{Q}_{\bv}(i) \bigg) \bigg)
\cdot \exp \bigg\{ -s \sum_{i=1}^{\ICC} \hat{Q}_{\bv}(i) \log \hat{p}_i \bigg\}.
\end{align}

It follows from the definition of $\hat{Q}_{\bv}(i)$ in \eqref{Def_Q_Alternative} that $s\hat{Q}_{\bv}(i) \in \{0,1,\ldots,s\}$ for any $i \in [\ICC]$. In order to bound the Gamma function factors in \eqref{ref2}, we invoke the inequality \cite{mortici2010sharp}  
\begin{align}
\label{ref3}
\Gamma(1+x) 
\leq \omega \sqrt{2\pi\bigg(x+\frac{1}{6}\bigg)} \bigg(\frac{x}{e}\bigg)^{x}
\end{align}
which holds for every $x \geq 1$, where $\omega = e\sqrt{\frac{3}{7\pi}}$. It can be  checked that \eqref{ref3} also holds at $x=0$. 

The inequality in \eqref{ref3} yields
\begin{align}
\Gamma\bigg(1+s\hat{Q}_{\bv}(i)\bigg)
&\leq \omega\sqrt{2\pi} (s\hat{Q}_{\bv}(i))^{s\hat{Q}_{\bv}(i)}e^{-s\hat{Q}_{\bv}(i)} \sqrt{s\hat{Q}_{\bv}(i)+\frac{1}{6}},
\end{align}
and in turn,
\begin{align}
\prod_{i=1}^{\ICC} \Gamma\bigg(1+s\hat{Q}_{\bv}(i)\bigg)
&\leq \prod_{i=1}^{\ICC} \omega\sqrt{2\pi} (s\hat{Q}_{\bv}(i))^{s\hat{Q}_{\bv}(i)}e^{-s\hat{Q}_{\bv}(i)} \sqrt{s\hat{Q}_{\bv}(i)+\frac{1}{6}} \\
\label{ref23}
&= (\omega\sqrt{2\pi})^{\ICC} s^s e^{-s} \prod_{i=1}^{\ICC} \hat{Q}_{\bv}(i)^{s\hat{Q}_{\bv}(i)} \prod_{i=1}^{\ICC} \sqrt{s\hat{Q}_{\bv}(i)+\frac{1}{6}}.
\end{align}

Now, 
\begin{align}
\prod_{i=1}^{\ICC} \sqrt{s\hat{Q}_{\bv}(i)+\frac{1}{6}}
&= \exp\bigg\{\frac{1}{2}\sum_{i=1}^{\ICC}\log\bigg(s\hat{Q}_{\bv}(i)+\frac{1}{6}\bigg) \bigg\} \\
\label{ref21}
&\leq \exp\bigg\{\frac{\ICC}{2}\log\bigg(\frac{s}{\ICC}+1\bigg) \bigg\},
\end{align}
where \eqref{ref21} follows from Jensen's inequality and the concavity of the logarithmic function.

Before we proceed, we recall that the $\chi^2$-divergence between two PMFs $\{P(x)\}_{x\in\calX}$ and $\{Q(x)\}_{x\in\calX}$ is defined by 
\begin{equation}
\chi^2(P\|Q) = \sum_{x\in\calX} \frac{(P(x)-Q(x))^2}{Q(x)} = \sum_{x\in\calX} \frac{P(x)^2}{Q(x)} - 1. 
\end{equation}

Substituting \eqref{ref21} back into \eqref{ref23} and then into \eqref{ref2}, we arrive at
\begin{align}
&\P[D(\hat{\bQ}_{\bv}\|\hat{\bP}_m) \leq D(\hat{\bQ}_{\bv}\|\hat{\bp})] \nn \\
&\leq  
\bigg(\frac{M}{M-\ICC}\bigg)^{s} \cdot
\frac{\Gamma(\ICC)}{\Gamma(\ICC+s)} \cdot (\omega\sqrt{2\pi})^{\ICC} s^s e^{-s} \nn \\ &~~~~~~~~~~~~~~~~~~\times \prod_{i=1}^{\ICC} \hat{Q}_{\bv}(i)^{s\hat{Q}_{\bv}(i)} \cdot \bigg(\frac{s}{\ICC}+1\bigg)^{\frac{\ICC}{2}} \cdot \exp\bigg\{s \sum_{i=1}^{\ICC} \hat{Q}_{\bv}(i) \log \frac{1}{\hat{p}_i} \bigg\} \\
&= (\omega\sqrt{2\pi})^{\ICC} \cdot
\bigg(\frac{M}{M-\ICC}\bigg)^{s} \cdot
\frac{\Gamma(\ICC)}{\Gamma(\ICC+s)} \cdot s^s \cdot e^{-s} \cdot \bigg(\frac{s}{\ICC}+1\bigg)^{\frac{\ICC}{2}} \cdot \exp\bigg\{ s \sum_{i=1}^{n} \hat{Q}_{\bv}(i) \log \frac{\hat{Q}_{\bv}(i)}{\hat{p}_i} \bigg\} \\
&= (\omega\sqrt{2\pi})^{\ICC} \cdot
\bigg(\frac{M}{M-\ICC}\bigg)^{s} \cdot
\frac{\Gamma(\ICC)}{\Gamma(\ICC+s)} \cdot s^s \cdot e^{-s} \cdot \bigg(\frac{s}{\ICC}+1\bigg)^{\frac{\ICC}{2}} \cdot \exp\{s \cdot D(\hat{\bQ}_{\bv}\|\hat{\bp})\} \\
\label{ref11}
&\leq (\omega\sqrt{2\pi})^{\ICC} \cdot
\bigg(\frac{M}{M-\ICC}\bigg)^{s} \cdot
\frac{\Gamma(\ICC)}{\Gamma(\ICC+s)} \cdot s^s \cdot e^{-s} \cdot \bigg(\frac{s}{\ICC}+1\bigg)^{\frac{\ICC}{2}} \cdot \exp\{s \cdot \chi^2(\hat{\bQ}_{\bv}\|\hat{\bp})\} \\
\label{ref4}
&\dfn A(\ICC,M,s) \cdot \exp\{s \cdot \chi^2(\hat{\bQ}_{\bv}\|\hat{\bp})\},
\end{align}
where \eqref{ref11} follows from the fact that \cite[Theorem 5]{gibbs2002choosing}
\begin{align}
D(\hat{\bQ}_{\bv}\|\hat{\bp})
&\leq \chi^2(\hat{\bQ}_{\bv}\|\hat{\bp}).
\end{align}

Upper-bounding \eqref{ref00} with \eqref{ref4} yields
\begin{align}
\varepsilon_M(\hat{\bp},\bu,\bv)
&\leq \min \bigg\{ 1, \sum_{m=2}^{|\calC_M|} A(\ICC,M,s) \cdot \exp\{s \cdot \chi^2(\hat{\bQ}_{\bv}\|\hat{\bp})\} \bigg\} \\
\label{ref62}
&\leq \min \bigg\{ 1, |\calC_M| \cdot A(\ICC,M,s) \cdot \exp\{s \cdot \chi^2(\hat{\bQ}_{\bv}\|\hat{\bp})\} \bigg\}.
\end{align}

In Appendix \ref{appendix_inequality} we prove that 
\begin{align}\label{UpperBound_A}
A(\ICC,M,s) 
&\leq 2\sqrt{1+\NOS} \cdot \exp\{(1+2\xi)\ICC\} \cdot \bigg(\frac{\ICC}{\ICC+s}\bigg)^{\frac{\ICC}{2}} \\
\label{ref63}
&\dfn B(\ICC,\NOS,s),
\end{align}
which is monotonically decreasing in $s$.

Upper-bounding \eqref{ref62} with \eqref{ref63} yields that
\begin{align}
\varepsilon_M(\hat{\bp},\bu,\bv)
&\leq \min \bigg\{ 1, |\calC_M| \cdot B(\ICC,\NOS,s) \cdot \exp\{s \cdot \chi^2(\hat{\bQ}_{\bv}\|\hat{\bp})\} \bigg\} \\
&\leq \min \bigg\{ 1, |\calC_M| \cdot B(\ICC,\NOS,s) \cdot \exp\{\NOS \cdot \chi^2(\hat{\bQ}_{\bv}\|\hat{\bp})\} \bigg\},
\end{align}
which holds since
\begin{equation}
s = \sum_{i=1}^{\ICC}v_i \leq \sum_{i=1}^{\ICC}u_i = \NOS.
\end{equation}

Given the samples $\bU=\bu$, we take the expectation with respect to the random sequencing, which induces the random vector $\bV$. We take this expectation in two steps: first over $\bV$ given $S=s$, and then over $S$.
For any $s\in\{0,\ldots,\NOS\}$, define the set
\begin{equation}
\calA(\bu,s) = \bigg\{ (v_1,\ldots,v_\ICC) : \forall i\in[\ICC], v_i\in\{0,\ldots,u_i\},~ \sum_{i=1}^{\ICC}v_i=s \bigg\}.
\end{equation}
Now, 
\begin{align}
\varepsilon_M(\hat{\bp},\bu)
\label{ref71}
&\leq \sum_{s=0}^{\NOS} P_{S}(s) \sum_{\bv\in\calA(\bu,s)} P_{\bV|S}(\bv|s) \min \bigg\{ 1, |\calC_M| \cdot B(\ICC,\NOS,s) \cdot \exp\{\NOS \cdot \chi^2(\hat{\bQ}_{\bv}\|\hat{\bp})\} \bigg\}.
\end{align}

Given $\hat{\bp}$, $\bu$, and $s$, we split the set $\calA(\bu,s)$ into two complementary subsets; $\bv$ vectors for which $\chi^2(\hat{\bQ}_{\bv}\|\hat{\bp})$ is relatively small and $\bv$ vectors for which $\chi^2(\hat{\bQ}_{\bv}\|\hat{\bp})$ is relatively large. We make the following definition.  
Let $\{\Delta_n\}_{n=1}^{\infty}$ be a monotonically increasing sequence with $\lim_{n\to\infty} \Delta_n=\infty$, to be chosen later.
For a given $\hat{\bp}$, $\bu$, and $s$, define $\calF(\hat{\bp},\bu,s)$ by
\begin{align} 
\calF(\hat{\bp},\bu,s) 
\label{ref20}
&= \bigg\{\bv\in\calA(\bu,s) ~\bigg|~ 0 \le \chi^2(\hat{\bQ}_{\bv}\|\hat{\bp}) \leq \Delta_M \cdot \E\bigg[ \chi^2(\hat{\bQ}_{\bV}\|\hat{\bp}) \bigg] \bigg\}.
\end{align}

Conditioned on $\hat{\bp}$, $\bu$, and $s$, we calculate the expectation of $\chi^2(\hat{\bQ}_{\bV}\|\hat{\bp})$.
Given $\bV=\bv=(v_1,\ldots,v_\ICC)$, we have that
\begin{align}
\chi^2(\hat{\bQ}_{\bv}\|\hat{\bp}) 
&= \sum_{i=1}^{\ICC} \frac{\hat{Q}_{\bv}^2(i)}{\hat{p}_i} - 1 \\
&= \sum_{i=1}^{\ICC} \frac{v_i^2}{s^2\hat{p}_i} - 1, 
\end{align}
and then
\begin{align}
\E\bigg[\chi^2(\hat{\bQ}_{\bV}\|\hat{\bp})\bigg]
&= \E\bigg[\sum_{i=1}^{\ICC} \frac{V_i^2}{s^2\hat{p}_i} - 1\bigg] \\
\label{ref70}
&= \sum_{i=1}^{\ICC} \frac{\E[V_i^2]}{s^2\hat{p}_i} - 1. 
\end{align}

In order to calculate the expectation in \eqref{ref70}, let us recall the following fact. Let $X \sim \text{Bin}(n,q)$ and $Y \sim \text{Bin}(m,q)$ be two independent random variables. Then, the conditional PMF of $X$ given $X+Y=s$ is given by the hypergeometric distribution:
\begin{equation}
    P_{X|X+Y=s}(k) = \frac{\binom{n}{k}\binom{m}{s-k}}{\binom{n+m}{s}}.
\end{equation}

For a hypergeometric random variable $Z$ with these parameters, it is known that 
\begin{align}
    \E[Z] &= \frac{ns}{m+n}, \\
    \text{Var}[Z] &= \frac{mns(m+n-s)}{(m+n)^2(m+n-1)},
\end{align}
and thus
\begin{align}
\E[Z^2] 
&= \text{Var}[Z] + (\E[Z])^2 \\
&= \frac{mns(m+n-s)}{(m+n)^2(m+n-1)} + \frac{n^2s^2}{(m+n)^2}.
\end{align}
In our setting, the relevant parameters are $n=u_i$ and $m=\NOS-u_i$.
Substituting these parameters yields 
\begin{align}
\E[V_i^2]
= \frac{u_i(\NOS-u_i)s(\NOS-s)}{\NOS^2(\NOS-1)} + \frac{u_i^2s^2}{\NOS^2},
\end{align}
which implies that
\begin{align}
\E\bigg[\chi^2(\hat{\bQ}_{\bV}\|\hat{\bp})\bigg]
&= \sum_{i=1}^{\ICC} \frac{1}{s^2\hat{p}_i} \cdot \bigg[\frac{u_i(\NOS-u_i)s(\NOS-s)}{\NOS^2(\NOS-1)} + \frac{u_i^2s^2}{\NOS^2}\bigg] - 1 \\
&= \sum_{i=1}^{\ICC} \bigg[\frac{u_i(\NOS-u_i)(\NOS-s)}{\NOS^2(\NOS-1)s\hat{p}_i} + \frac{u_i^2}{\NOS^2\hat{p}_i}\bigg] - 1 \\
&\dfn F(\hat{\bp},\bu,s),
\end{align}
which is monotonically decreasing in $s$.

The inner sum in \eqref{ref71} is calculated as follows:
\begin{align}
&\sum_{\bv\in\calA(\bu,s)} P_{\bV|S}(\bv|s) \min \bigg\{ 1, |\calC_M| \cdot B(\ICC,\NOS,s) \cdot \exp\{\NOS \cdot \chi^2(\hat{\bQ}_{\bv}\|\hat{\bp})\} \bigg\} \nn \\
&= \sum_{\bv\in\calF} P_{\bV|S}(\bv|s) \min \bigg\{ 1, |\calC_M| \cdot B(\ICC,\NOS,s) \cdot \exp\{\NOS \cdot \chi^2(\hat{\bQ}_{\bv}\|\hat{\bp})\} \bigg\} \nn \\
&~~~~+\sum_{\bv\in\calF^{\mbox{\footnotesize c}}} P_{\bV|S}(\bv|s) \min \bigg\{ 1, |\calC_M| \cdot B(\ICC,\NOS,s) \cdot \exp\{\NOS \cdot \chi^2(\hat{\bQ}_{\bv}\|\hat{\bp})\} \bigg\} \\
\label{ref72}
&\leq \sum_{\bv\in\calF} P_{\bV|S}(\bv|s) \min \bigg\{ 1, |\calC_M| \cdot B(\ICC,\NOS,s) \cdot \exp\{\NOS \cdot \Delta_M \cdot F(\hat{\bp},\bu,s)\} \bigg\} + \sum_{\bv\in\calF^{\mbox{\footnotesize c}}} P_{\bV|S}(\bv|s)  \\
&= \P[\bV \in \calF] \cdot \min \bigg\{ 1, |\calC_M| \cdot B(\ICC,\NOS,s) \cdot \exp\{\NOS \cdot \Delta_M \cdot F(\hat{\bp},\bu,s)\} \bigg\} + \P\bigg[\bV \in \calF^{\mbox{\footnotesize c}} \bigg]  \\
\label{ref73}
&\leq  \min \bigg\{ 1, |\calC_M| \cdot B(\ICC,\NOS,s) \cdot \exp\{\NOS \cdot \Delta_M \cdot F(\hat{\bp},\bu,s)\} \bigg\} + \P\bigg[\bV \in \calF^{\mbox{\footnotesize c}} \bigg],
\end{align}
where \eqref{ref72} follows since for any $\bv \in \calF$, $\chi^2(\hat{\bQ}_{\bv}\|\hat{\bp})$ is upper-bounded by $\Delta_M \cdot F(\hat{\bp},\bu,s)$, and the right-hand side summation is bounded using $\min\{1,t\} \leq 1$.

It follows by Markov's inequality that 
\begin{align}
\P[\bV \in \calF^{\mbox{\footnotesize c}}]
&= \P\bigg[ \chi^2(\hat{\bQ}_{\bV}\|\hat{\bp}) \geq \Delta_M \cdot \E[\chi^2(\hat{\bQ}_{\bV}\|\hat{\bp})] \bigg] \\
\label{ref74}
&\leq \frac{1}{\Delta_M},
\end{align}
which converges to zero as $M \to \infty$ since we assume that $\{\Delta_n\}_{n=1}^{\infty}$ is a monotonically increasing sequence with $\lim_{n\to\infty} \Delta_n=\infty$.

Upper-bounding \eqref{ref71} with \eqref{ref73} and \eqref{ref74} yields  
\begin{align}
\varepsilon_M(\hat{\bp},\bu)
&\leq \sum_{s=0}^{\NOS} P_{S}(s) \bigg[ \min \bigg\{ 1, |\calC_M| \cdot B(\ICC,\NOS,s) \cdot \exp\{\NOS \cdot \Delta_M \cdot F(\hat{\bp},\bu,s)\} \bigg\} + \frac{1}{\Delta_M} \bigg] \\
&= \sum_{s=0}^{\NOS} P_{S}(s) \cdot \min \bigg\{ 1, |\calC_M| \cdot B(\ICC,\NOS,s) \cdot \exp\{\NOS \cdot \Delta_M \cdot F(\hat{\bp},\bu,s)\} \bigg\} + \frac{1}{\Delta_M}.
\end{align}

Denote $P_{\mbox{\scriptsize c}}(\calC^{\star})=1-P_{\mbox{\scriptsize er}}(\calC^{\star})$, and observe that
\begin{equation}
    S = \sum_{i=1}^{\ICC} V_i \sim \text{Bin}\bigg(\sum_{i=1}^{\ICC} u_i, P_{\mbox{\scriptsize c}}(\calC^{\star}) \bigg)
  = \text{Bin}\bigg(\NOS,P_{\mbox{\scriptsize c}}(\calC^{\star}) \bigg).  
\end{equation}

For a given $\kappa \in (0,1)$, define 
\begin{equation}
s^{*} = \NOS P_{\mbox{\scriptsize c}}(\calC^{\star})(1-\kappa) \dfn \eta \NOS,
\end{equation}
and then
\begin{align}
&\varepsilon_M(\hat{\bp},\bu) \nn \\
&\leq \sum_{s=0}^{s^{*}} P_{S}(s) \cdot \min \bigg\{ 1, |\calC_M| \cdot B(\ICC,\NOS,s) \cdot \exp\{\NOS \cdot \Delta_M \cdot F(\hat{\bp},\bu,s)\} \bigg\} \nn \\
&~~+\sum_{s=s^{*}}^{\NOS} P_{S}(s) \cdot \min \bigg\{ 1, |\calC_M| \cdot B(\ICC,\NOS,s) \cdot \exp\{\NOS \cdot \Delta_M \cdot F(\hat{\bp},\bu,s)\} \bigg\} + \frac{1}{\Delta_M} \\
\label{ref75}
&\leq \sum_{s=0}^{s^{*}} P_{S}(s) +\sum_{s=s^{*}}^{\NOS} P_{S}(s) \cdot \min \bigg\{ 1, |\calC_M| \cdot B(\ICC,\NOS,\eta \NOS) \cdot \exp\bigg\{ \NOS \cdot \Delta_M \cdot F(\hat{\bp},\bu,\eta \NOS) \bigg\} \bigg\} + \frac{1}{\Delta_M} \\
&= \P[S \leq s^{*}] + \P[S \geq s^{*}] \cdot \min \bigg\{ 1, |\calC_M| \cdot B(\ICC,\NOS,\eta \NOS) \cdot \exp\bigg\{ \NOS \cdot \Delta_M \cdot F(\hat{\bp},\bu,\eta \NOS) \bigg\} \bigg\} + \frac{1}{\Delta_M} \\
\label{ref76}
&\leq \P[S \leq s^{*}] + \min \bigg\{ 1, |\calC_M| \cdot B(\ICC,\NOS,\eta \NOS) \cdot \exp\bigg\{ \NOS \cdot \Delta_M \cdot F(\hat{\bp},\bu,\eta \NOS) \bigg\} \bigg\} + \frac{1}{\Delta_M},
\end{align}
where \eqref{ref75} follows from the monotone decrease in $s$.

The probability in \eqref{ref76} is bounded as
\begin{align}
\P[S \leq s^{*}]
&= \P\bigg[ S \leq \NOS P_{\mbox{\scriptsize c}}(\calC^{\star})(1-\kappa) \bigg] \\
\label{ref77}
&\leq \exp\bigg\{-\frac{1}{2}\kappa^2 \NOS  P_{\mbox{\scriptsize c}}(\calC^{\star}) \bigg\},
\end{align}
where \eqref{ref77} follows from the multiplicative Chernoff bound.

We upper-bound the quantity $B(\ICC,\NOS,\eta \NOS)$ as follows:
\begin{align}
B(\ICC,\NOS,\eta \NOS)
&= 2\sqrt{1+\NOS} \cdot \exp\{(1+2\xi)\ICC\} \cdot \bigg(\frac{\ICC}{\ICC+\eta \NOS}\bigg)^{\frac{\ICC}{2}}\\
\label{ref60}
&\leq 2\sqrt{1+\xi M} \cdot \exp\{(1+2\xi)\ICC\} \cdot \bigg(\frac{\ICC}{\eta\xi M}\bigg)^{\frac{\ICC}{2}}\\
\label{ref61}
&= 2\sqrt{1+\xi M} \cdot \exp\bigg\{\bigg(1+2\xi-\frac{1}{2}\log(\eta\xi) \bigg)\ICC \bigg\} \cdot \bigg(\frac{\ICC}{M}\bigg)^{\frac{\ICC}{2}} \\
\label{expression_C}
&\dfn 2\sqrt{1+\xi M} \cdot \exp\{\varphi(\xi,\eta)\ICC\} \cdot \bigg(\frac{\ICC}{M}\bigg)^{\frac{\ICC}{2}} \\
&\dfn C(\ICC,M),
\end{align}
where \eqref{ref60} follows since $\NOS=\xi M$. 

The quantity $F(\hat{\bp},\bu,\eta \NOS)$ is given by
\begin{align}
F(\hat{\bp},\bu,\eta \NOS)
&= \sum_{i=1}^{\ICC} \bigg[\frac{u_i(\NOS-u_i)(\NOS-\eta \NOS)}{\NOS^2(\NOS-1)\eta \NOS \hat{p}_i} + \frac{u_i^2}{\NOS^2\hat{p}_i}\bigg] - 1 \\
&= \sum_{i=1}^{\ICC} \bigg[\frac{\chi(u_i \NOS-u_i^2)}{\NOS^2(\NOS-1)\hat{p}_i} + \frac{u_i^2}{\NOS^2\hat{p}_i}\bigg] - 1 \\
&\dfn G(\hat{\bp},\bu),
\end{align}
where we have denoted $\chi=\frac{1-\eta}{\eta}$.

Putting everything together, we have that
\begin{align}
&\varepsilon_M(\hat{\bp},\bu) \nn \\
&\leq \exp\bigg\{-\frac{1}{2}\kappa^2 \NOS P_{\mbox{\scriptsize c}}(\calC^{\star}) \bigg\} + \min \bigg\{ 1, |\calC_M| \cdot C(\ICC,M) \cdot \exp\bigg\{ \NOS \cdot \Delta_M \cdot G(\hat{\bp},\bu) \bigg\} \bigg\} + \frac{1}{\Delta_M} \\
&\dfn \exp\bigg\{-\frac{1}{2}\kappa^2 \NOS P_{\mbox{\scriptsize c}}(\calC^{\star}) \bigg\} 
+ \Psi(\hat{\bp},\bu) + \frac{1}{\Delta_M}.
\end{align}

Given $\hat{\bp}$, we take the expectation with respect to the random sampling. 
We split the space of $\bu$ vectors into two complementary subsets; $\bu$ vectors for which $G(\hat{\bp},\bu)$ is relatively small and $\bu$ vectors for which $G(\hat{\bp},\bu)$ is relatively large. We make the following definition.  
For a given $\hat{\bp}$, define $\calG(\hat{\bp})$ by
\begin{align} 
\calG(\hat{\bp}) 
\label{ref20}
&= \bigg\{\bu ~\bigg|~ G(\hat{\bp},\bu) \leq \Delta_M \cdot \E[G(\hat{\bp},\bU)] \bigg\}.
\end{align}

The expectation in \eqref{ref20} is calculated as follows:
\begin{align}
\E[G(\hat{\bp},\bU)]  
&= \sum_{i=1}^{\ICC} \bigg[\frac{\chi(\E[U_i]\NOS-\E[U_i^2])}{\NOS^2(\NOS-1)\hat{p}_i} + \frac{\E[U_i^2]}{\NOS^2\hat{p}_i}\bigg] - 1 \\
\label{ref24}
&= \sum_{i=1}^{\ICC} \bigg[\frac{\chi(\hat{p}_i \NOS^2-\NOS\hat{p}_i(1-\hat{p}_i) - \NOS^2\hat{p}_i^2)}{\NOS^2(\NOS-1)\hat{p}_i} + \frac{\NOS\hat{p}_i(1-\hat{p}_i) + \NOS^2\hat{p}_i^2}{\NOS^2\hat{p}_i}\bigg] - 1 \\
&= \sum_{i=1}^{\ICC} \bigg[\frac{\chi[\NOS^2\hat{p}_i(1-\hat{p}_i) -\NOS\hat{p}_i(1-\hat{p}_i)]}{\NOS^2(\NOS-1)\hat{p}_i} + \frac{\NOS\hat{p}_i(1-\hat{p}_i) + \NOS^2\hat{p}_i^2}{\NOS^2\hat{p}_i}\bigg] - 1 \\
&= \sum_{i=1}^{\ICC} \bigg[\frac{\chi \NOS(\NOS-1)\hat{p}_i(1-\hat{p}_i)}{\NOS^2(\NOS-1)\hat{p}_i} + \frac{\NOS\hat{p}_i(1-\hat{p}_i) + \NOS^2\hat{p}_i^2}{\NOS^2\hat{p}_i}\bigg] - 1 \\
&= \sum_{i=1}^{\ICC} \bigg[\frac{\chi(1-\hat{p}_i)}{\NOS} + \frac{1-\hat{p}_i + \NOS\hat{p}_i}{\NOS}\bigg] - 1 \\
&= \frac{\chi(\ICC-1)}{\NOS} + \frac{\ICC-1+\NOS}{\NOS} - 1 \\
&= \frac{\chi(\ICC-1)}{\NOS} + \frac{\ICC-1}{\NOS}  \\
&= \frac{(1+\chi)(\ICC-1)}{\NOS}  \\
&\dfn \Phi(\ICC,\NOS,\chi),
\end{align}
where \eqref{ref24} follows by expanding the second moment of a binomial random variable.

Averaging with respect to the random sampling, we obtain
\begin{align}
\E[\Psi(\hat{\bp},\bU)]
&= \sum_{\bu} P_{\bU}(\bu) \Psi(\hat{\bp},\bu) \\
&= \sum_{\bu} P_{\bU}(\bu) \min \bigg\{ 1, |\calC_M| \cdot C(\ICC,M) \cdot \exp\bigg\{\NOS \cdot \Delta_M \cdot G(\hat{\bp},\bu) \bigg\} \bigg\} \\
&= \sum_{\bu \in \calG} P_{\bU}(\bu) \min \bigg\{ 1, |\calC_M| \cdot C(\ICC,M) \cdot \exp\bigg\{\NOS \cdot \Delta_M \cdot G(\hat{\bp},\bu) \bigg\} \bigg\} \nn \\
&~~~~~~+\sum_{\bu \in \calG^{\mbox{\footnotesize c}}} P_{\bU}(\bu) \min \bigg\{ 1, |\calC_M| \cdot C(\ICC,M) \cdot \exp\bigg\{\NOS \cdot \Delta_M \cdot G(\hat{\bp},\bu) \bigg\} \bigg\} \\
\label{ref25}
&\leq \sum_{\bu \in \calG} P_{\bU}(\bu) \min \bigg\{ 1, |\calC_M| \cdot C(\ICC,M) \cdot \exp\bigg\{ \NOS \cdot \Delta_M^2 \cdot \Phi(\ICC,\NOS,\chi) \bigg\} \bigg\} + \sum_{\bu \in \calG^{\mbox{\footnotesize c}}} P_{\bU}(\bu)  \\
&= \P[\bU \in \calG] \cdot \min \bigg\{ 1, |\calC_M| \cdot C(\ICC,M) \cdot \exp\{\Delta_M^2(1+\chi)(\ICC-1)\} \bigg\} + \P[\bU \in \calG^{\mbox{\footnotesize c}}] \\
\label{ref30}
&\leq \min\bigg\{1,|\calC_M| \cdot C(\ICC,M) \cdot \exp\{\Delta_M^2(1+\chi)\ICC\} \bigg\} + \P[\bU \in \calG^{\mbox{\footnotesize c}}],
\end{align}
where \eqref{ref25} follows since for any $\bu \in \calG$, $G(\hat{\bp},\bu)$ is upper-bounded by $\Delta_M \cdot \Phi(\ICC,\NOS,\chi)$, and the right-hand side summation is bounded using $\min\{1,t\} \leq 1$.

It follows by Markov's inequality that 
\begin{align}
\P[\bU \in \calG^{\mbox{\footnotesize c}}]
&= \P\bigg[ G(\hat{\bp},\bU) \geq \Delta_M \cdot \E[G(\hat{\bp},\bU)] \bigg] \\
&\leq \frac{1}{\Delta_M},
\end{align}
which converges to zero as $M \to \infty$ since we assume that $\{\Delta_n\}_{n=1}^{\infty}$ is a monotonically increasing sequence with $\lim_{n\to\infty} \Delta_n=\infty$.

We continue by upper-bounding the expression in \eqref{ref30}: 
\begin{align}
\E[\Psi(\hat{\bp},\bU)]
&\leq \min\bigg\{1,|\calC_M| \cdot C(\ICC,M) \cdot \exp\{\Delta_M^2(1+\chi)\ICC\} \bigg\} + \frac{1}{\Delta_M} \\
&\leq |\calC_M| \cdot C(\ICC,M) \cdot \exp\{\Delta_M^2(1+\chi)\ICC\} + \frac{1}{\Delta_M},
\end{align}
which implies that
\begin{align}
\varepsilon_M(\hat{\bp}) 
&\leq \exp\bigg\{-\frac{1}{2}\kappa^2 \NOS P_{\mbox{\scriptsize c}}(\calC^{\star}) \bigg\} 
+ \E[\Psi(\hat{\bp},\bU)] + \frac{1}{\Delta_M} \\
\label{ref79}
&\leq \exp\bigg\{-\frac{1}{2}\kappa^2 \NOS P_{\mbox{\scriptsize c}}(\calC^{\star}) \bigg\} 
+ |\calC_M| \cdot C(\ICC,M) \cdot \exp\{\Delta_M^2(1+\chi)\ICC\} + \frac{2}{\Delta_M}.
\end{align}
The upper bound in \eqref{ref79} is independent of the realization of $\bP_1$, and hence 
\begin{align}
\varepsilon_M
\label{ref78}
&\leq \exp\bigg\{-\frac{1}{2}\kappa^2\xi M P_{\mbox{\scriptsize c}}(\calC^{\star}) \bigg\} 
+ |\calC_M| \cdot C(\ICC,M) \cdot \exp\{\Delta_M^2(1+\chi)\ICC\} + \frac{2}{\Delta_M},
\end{align}
where \eqref{ref78} follows since $\NOS=\xi M$.

Substituting the expression for $C(\ICC,M)$ from \eqref{expression_C}, we find that
\begin{align}
\varepsilon_M
&\leq \exp\bigg\{-\frac{1}{2}\kappa^2\xi M P_{\mbox{\scriptsize c}}(\calC^{\star}) \bigg\} + |\calC_M| \cdot 2\sqrt{1+\xi M} \cdot \exp\{\varphi(\xi,\eta)\ICC\} \nn \\
&~~~~~~~~~~\times \exp\bigg\{-\frac{\ICC}{2}\log\bigg(\frac{M}{\ICC}\bigg) \bigg\} \cdot \exp\{\Delta_M^2(1+\chi)\ICC\} + \frac{2}{\Delta_M}.
\end{align}

For some $\sigma > 0$, let the codebook size be
\begin{equation}
|\calC_M| = \exp\bigg\{\bigg(\frac{1}{2}-\sigma\bigg) \ICC \log\bigg(\frac{M}{\ICC}\bigg)\bigg\},
\end{equation}
which implies that for all $M$ sufficiently large
\begin{align}
\varepsilon_M
&\leq \exp\bigg\{-\frac{1}{2}\kappa^2\xi M P_{\mbox{\scriptsize c}}(\calC^{\star}) \bigg\} + 2\sqrt{1+\xi M} \cdot \exp\{\varphi(\xi,\eta)\ICC\} \nn \\
&~~~~~~~~~~\times \exp\bigg\{-\sigma \ICC \log\bigg(\frac{M}{\ICC}\bigg) \bigg\} \cdot \exp\{\Delta_M^2(1+\chi)\ICC\} + \frac{2}{\Delta_M}.
\end{align}

Finally, choosing $\Delta_M = \log^{\nu}(M)$ for some $\nu\in(0,\frac{1}{2})$, substituting $\ICC=M^{\gamma}$ with $\gamma=\beta R$, as well as $\chi=\frac{1-\eta}{\eta}$ and $\eta=P_{\mbox{\scriptsize c}}(\calC^{\star})(1-\kappa)$, yields that for all $M$ sufficiently large
\begin{align} 
\varepsilon_M
\label{ref_final}
&\leq 2\sqrt{1+\xi M} \cdot \exp\bigg\{ \bigg[\varphi(\xi,\eta) + \frac{\log^{2\nu}(M)}{P_{\mbox{\scriptsize c}}(\calC^{\star})(1-\kappa)}  - \sigma (1-\gamma)\log(M) \bigg] \cdot M^{\gamma} \bigg\} \nn \\
&~~~~~~~~~~+ \frac{2}{\log^{\nu}(M)} + \exp\bigg\{-\frac{1}{2}\kappa^2\xi M P_{\mbox{\scriptsize c}}(\calC^{\star}) \bigg\}.
\end{align}

Recall from \eqref{ref61} that
\begin{align}
\varphi(\xi,\eta)
&= 1+2\xi-\frac{1}{2}\log(\eta\xi) \\
&= 1+2\xi-\frac{1}{2}\log[P_{\mbox{\scriptsize c}}(\calC^{\star})(1-\kappa)\xi].
\end{align}

For any fixed $R\in(0,R_{\mbox{\scriptsize max}}(W))$, it follows that $P_{\mbox{\scriptsize c}}(\calC^{\star}) \to 1$ as $M \to \infty$, and then, the expression 
\begin{align}
&\varphi(\xi,\eta) + \frac{\log^{2\nu}(M)}{P_{\mbox{\scriptsize c}}(\calC^{\star})(1-\kappa)}  - \sigma (1-\gamma)\log(M) \nn \\
\label{Final_Expression}
&~~=1+2\xi-\frac{1}{2}\log[P_{\mbox{\scriptsize c}}(\calC^{\star})(1-\kappa)\xi] + \frac{\log^{2\nu}(M)}{P_{\mbox{\scriptsize c}}(\calC^{\star})(1-\kappa)}  - \sigma (1-\gamma)\log(M)
\end{align}
converges to $-\infty$ as $M \to \infty$ for any $\xi>0$, $\sigma >0$, $\gamma\in(0,1)$, $\nu \in (0,\frac{1}{2})$, and $\kappa\in(0,1)$, and hence, the error probability bound in \eqref{ref_final} converges to zero as $M \to \infty$.   

Since $\sigma>0$ can be made arbitrarily small, this completes the proof of Theorem \ref{Thm_main}.

\section{Proof of \eqref{UpperBound_A}
\label{appendix_inequality}}

Recall that 
\begin{equation}
\label{ref6}
A(\ICC,M,s) = (\omega\sqrt{2\pi})^{\ICC} \cdot
\bigg(\frac{M}{M-\ICC}\bigg)^{s} \cdot
\frac{\Gamma(\ICC)}{\Gamma(\ICC+s)} \cdot s^s \cdot e^{-s} \cdot \bigg(\frac{s+\ICC}{\ICC}\bigg)^{\frac{\ICC}{2}}.
\end{equation}

The second factor in \eqref{ref6} is bounded as
\begin{align}
\bigg(\frac{M}{M-\ICC}\bigg)^{s}
&\leq \bigg(\frac{M}{M-\ICC}\bigg)^{\NOS} \\
&= \exp\bigg\{\NOS \log\bigg(\frac{M}{M-\ICC}\bigg) \bigg\} \\
&= \exp\bigg\{\NOS \log\bigg(1 + \frac{\ICC}{M-\ICC}\bigg) \bigg\} \\
\label{ref26}
&\leq \exp\bigg\{\frac{\ICC \NOS}{M-\ICC} \bigg\} \\
\label{ref7}
&\leq \exp\bigg\{\frac{\ICC \NOS}{M-\frac{M}{2}} \bigg\} \\
\label{ref9}
&= \exp\{2\xi \ICC\},
\end{align} 
where \eqref{ref26} is due to $\log(1+t) \leq t$, \eqref{ref7} follows because $\ICC = M^{\beta R}$ for some $\beta \in (0,\frac{1}{\log|\calX|})$, and since $R<R_{\mbox{\scriptsize max}}(W) \leq \log|\calX|$, it holds that $\ICC \leq \frac{M}{2}$ for all $M$ sufficiently large, and in \eqref{ref9} we used the definition $\xi=\frac{\NOS}{M}$.

We invoke the following double-sided inequality from \cite[Theorem 5]{gordon1994stochastic}. For any $t>0$,
\begin{align} \label{Gamma_Bounds}
\sqrt{2\pi}t^{t-1/2}e^{-t} \leq \Gamma(t) \leq \sqrt{2\pi}t^{t-1/2}e^{-t}e^{\frac{1}{12t}},
\end{align}
and thus, the third factor in \eqref{ref6} is bounded as follows.
\begin{align}
\frac{\Gamma(\ICC)}{\Gamma(\ICC+s)}    
&\leq \frac{\sqrt{2\pi}\ICC^{\ICC-1/2}e^{-\ICC}e^{\frac{1}{12\ICC}}}{\sqrt{2\pi}(\ICC+s)^{\ICC+s-1/2}e^{-(\ICC+s)}} \\
&= \sqrt{1+\frac{s}{\ICC}} \cdot \frac{\ICC^\ICC}{(\ICC+s)^{\ICC}} \cdot \frac{1}{(\ICC+s)^{s}} \cdot e^s e^{\frac{1}{12\ICC}} \\
\label{ref8}
&\leq 2\sqrt{1+\NOS} \cdot \bigg(\frac{\ICC}{\ICC+s}\bigg)^\ICC \cdot \frac{1}{(\ICC+s)^{s}} \cdot e^s,
\end{align}
where in \eqref{ref8} we upper-bounded $e^{\frac{1}{12\ICC}} \leq 2$, which holds for any $\ICC \in \{1,2,\ldots\}$.

Upper-bounding \eqref{ref6} with \eqref{ref9} and \eqref{ref8} yields that for all $M$ sufficiently large
\begin{align}
&A(\ICC,M,s) \nn \\
&= (\omega\sqrt{2\pi})^{\ICC} \cdot
\bigg(\frac{M}{M-\ICC}\bigg)^{s} \cdot
\frac{\Gamma(\ICC)}{\Gamma(\ICC+s)} \cdot s^s \cdot e^{-s} \cdot \bigg(\frac{s+\ICC}{\ICC}\bigg)^{\frac{\ICC}{2}} \\
&\leq (\omega\sqrt{2\pi})^{\ICC} \cdot
\exp\{2\xi \ICC\} \cdot
2\sqrt{1+\NOS} \cdot \bigg(\frac{\ICC}{\ICC+s}\bigg)^\ICC \cdot \frac{1}{(\ICC+s)^{s}} \cdot e^s \cdot s^s \cdot e^{-s} \cdot \bigg(\frac{s+\ICC}{\ICC}\bigg)^{\frac{\ICC}{2}} \\
&= 2\sqrt{1+\NOS} \cdot (\omega\sqrt{2\pi})^{\ICC} \cdot
\exp\{2\xi \ICC\} \cdot \frac{s^s}{(\ICC+s)^{s}} \cdot \bigg(\frac{\ICC}{\ICC+s}\bigg)^{\frac{\ICC}{2}} \\
\label{ref27}
&\leq 2\sqrt{1+\NOS} \cdot
\exp\bigg\{ \ICC \log\bigg(e\sqrt{\frac{6}{7}}\bigg) \bigg\} \cdot \exp\{2\xi \ICC\} \cdot \bigg(\frac{\ICC}{\ICC+s}\bigg)^{\frac{\ICC}{2}} \\
&\leq 2\sqrt{1+\NOS} \cdot \exp\{(1+2\xi)\ICC\} \cdot \bigg(\frac{\ICC}{\ICC+s}\bigg)^{\frac{\ICC}{2}},
\end{align}
where \eqref{ref27} follows by substituting $\omega = e\sqrt{\frac{3}{7\pi}}$ and since $s^s/(\ICC+s)^s \leq 1$.

\bibliographystyle{IEEEtran}
\bibliography{DNA_short_noisy_v01.bib}






\end{document}